\documentclass[aps,prd,preprint]{revtex4}
\usepackage{epsfig}
\usepackage{amssymb}

\usepackage{graphicx,psfrag}
\usepackage{amsmath}
\usepackage{bm}

\newcommand{\beqa}{\begin{eqnarray}}
\newcommand{\eeqa}{\end{eqnarray}}
\newcommand{\beq}{\begin{equation}}
\newcommand{\eeq}{\end{equation}}
\newcommand{\bsp}{\begin{split}}
\newcommand{\esp}{\end{split}}
\newcommand{\bal}{\begin{align}}
\newcommand{\eal}{\end{align}}

\begin{document}

\title{Operator analysis of effective spin-flavor interactions for $L=1$ excited baryons}

\def\addCS{Departamento de F\'{\i}sica, FCEyN, Universidad de Buenos Aires and IFIBA, 
CONICET, Ciudad Universitaria, Pabell\'on~1, (1428) Buenos Aires, Argentina}

\author{Cintia Willemyns}\affiliation{\addCS}
\author{Carlos Schat}\affiliation{\addCS}

\begin{abstract} \vspace*{18pt}

We match  the  non-relativistic quark model,  with both flavor dependent and
flavor independent effective quark-quark interactions, to the spin-flavor
operator basis of the $1/N_c$ expansion for the $L=1$ non-strange baryons.  
We obtain analytic expressions for
the coefficients of the $1/N_c$ operators in terms of radial integrals that
depend on the shape and relative strength of the spin-spin, spin-orbit and
tensor interactions of the model, which are left unspecified. We obtain several new,
parameter-free relations between the seven masses and the two mixing angles that
can discriminate between different spin-flavor structures of the effective
quark-quark interaction. We discuss in detail how a general parametrization of 
the mass matrix
depends on the mixing angles and is constrained by the assumptions on the
effective quark-quark interaction. We find that, within the
present experimental uncertainties, consistency with the best values of the
mixing angles as determined by a recent global fit to masses and decays does
not exclude any of the two most extreme possibilities of flavor dependent
(independent) quark-quark interactions, as generated by meson (gluon) exchange
interactions.  

\end{abstract} 


\maketitle

\section{Introduction}

In the last 15 years a large amount of data on electroproduction of mesons was
accumulated at different facilities, with the purpose of determining the
resonance contribution to the cross sections and identifying the members of the
excited baryon spectrum. The goal is to obtain a precise description of the
first excited states of the ground state baryons, which is an important test
for our understanding of QCD in the low-energy, strong-coupling regime. Models
try to capture the most relevant physics giving a description in terms of
effective degrees of freedom, but it is hard to make an estimate of the
theoretical errors involved and there is little guide on how to improve on
them. On the other hand, lattice calculations are based on first principles and
recently they have shown a significant progress in the determination of the light
baryon mass spectrum and the identification of the spin-parity of the excited
states \cite{Edwards:2011jj,Edwards:2012fx}. The lattice results seem to
confirm the old quark model picture, where the light-flavored baryons fill
irreducible representations of the orbital $\times$ spin-flavor group
$O(3)\times SU(6)$.
Empirically there are still many missing states in the spin-flavor multiplets of the 
quark model, an old
problem that has constantly spurred the experimental program
in the search of new resonances.  For recent reviews on the experimental status
of excited baryons and quark model related discussions see \cite{Capstick:2000qj}
\cite{Klempt:2009pi}
\cite{Crede:2013kia}.

An alternative approach to study the phenomenology of baryons is to 
consider 
the large $N_c$ limit of QCD \cite{'tHooft:1973jz} \cite{Witten:1979kh}, where 
it has been shown that the
spin-flavor symmetry for baryons arises from consistency relations for
pion-nucleon scattering  \cite{Gervais:1983wq} \cite{Gervais:1984rc}
\cite{Bardakci:1983ev} \cite{Dashen:1993as} \cite{Jenkins:1993zu}. The
predictions of this symmetry for the masses and the couplings explain some of
the successes of the non-relativistic quark model \cite{Dashen:1993jt} . The
breaking of the spin-flavor symmetry can be studied systematically in a $1/N_c$
expansion  using quark operators, establishing a deep connection between QCD and
the quark model \cite{Carone:1993dz} \cite{Luty:1993fu} \cite{Dashen:1994qi}.

The $1/N_c$ expansion was first applied to the masses and axial couplings of
ground state baryons \cite{Carone:1993dz} \cite{Luty:1993fu}
\cite{Dashen:1993jt} \cite{Dashen:1994qi} and later to study the masses of the
negative parity $L=1$ excited baryons \cite{Goity:1996hk} \cite{Pirjol:1997bp}
\cite{Pirjol:1997sr} \cite{Carlson:1998vx} \cite{Schat:2001xr}
\cite{Goity:2002pu} with great success. The strong and electromagnetic
properties of these states, as well as of resonances in other spin-flavor
multiplets have also been studied in the $1/N_c$ expansion and a
number of interesting results were obtained, see \cite{Matagne:2014lla} for a review, and
references therein.

Recently, the  $1/N_c$ expansion has also been put in connection with the
lattice calculations of the baryon spectrum \cite{Jenkins:2009wv}
\cite{Cordon:2014sda} \cite{Fernando:2014dna} discussing the spin-flavor
composition of the baryons and the quark mass dependence of the results.
However, in the systematic operator expansion provided by the  $1/N_c$
approach, all the non-perturbative dynamics remains hidden in the values of the
operator coefficients that in a phenomenological analysis have to be fitted to
data.  The physics driving the relative numerical sizes of  operator
coefficients remains unknown. Comparison with the results of lattice
calculations give some insight on their quark mass dependence but an
interpretation in terms of effective degrees of freedom and their interactions
is missing.

It is therefore very desirable to put the $1/N_c$ expansion in closer connection with 
dynamical calculations, such as the one provided by the quark model, in order to arrive  
at a simple physical picture in terms of effective
degrees of freedom. 
As a first step toward this goal, a very general method to match quark model
interactions to the $1/N_c$ operators was presented in
Ref.~\cite{Pirjol:2007ed} using the permutation group $S_N$, for arbitrary
$N_c$, generalizing a similar analysis for $N_c=3$ done by Collins and Georgi
in Ref.~\cite{Collins:1998ny}.  As an application, the most general two-body
interaction excluding three-body forces was considered in \cite{Pirjol:2008gd} 
without making any assumptions on the form of the confining potential, obtaining 
as a result Eq.~(5) in Ref.~\cite{Pirjol:2008gd}, which gives 
a correlation between the mixing angles of the $L=1$ non-strange
excited baryons. 

In Ref.~\cite{Galeta:2009pn}, a particular version of the quark model (QM) containing
only spin-spin and quadrupole flavor-independent interactions,  
the Isgur-Karl (IK) model \cite{Isgur:1977ef} \cite{Isgur:1978xj}, has been matched
to the $1/N_c$ operator expansion. Due to the harmonic oscillator confining
potential adopted in this model, the wave functions can be found exactly, and
the center of mass motion can be taken into account
explicitly. This analysis was extended in Ref.~\cite{Pirjol:2010th}   by keeping only the spin-spin
interaction but allowing for the most general flavor dependence of the
quark-quark interaction, neglecting the quadrupole and spin-orbit
interaction. In the present work we consider the most general quark-quark spin-flavor
interaction, taking into account all three components of the interaction, and
allowing for the most general spin-flavor structure.

We focus here on the mass spectrum of the lightest $L=1$  baryons, the
nonstrange members of the negative parity SU(6) $\bf 70$-plet. 
Configuration mixing was neglected as has been done in previous $1/N_c$
analysis, noting that this mixing is driven by the numerically suppressed
spin-orbit operators \cite{Fernando:2014dna}.  

The  general spin-flavor structure of the interactions can be reduced to two
extreme cases, where we only have flavor independent interactions as obtained in
model interactions based on one-gluon exchange (OGE), or flavor dependent
interactions that arise in model interactions based on one-meson exchange
(OME). 
Considering different spin-flavor structures for the interaction we obtain
several new mass-angle relations that are parameter free. Even in the most general
case they are sufficient to constrain the mixing angles. They will be shown to
be compatible with the latest determination of the angles from a global fit to 
the masses, decay widths and 
photo couplings~\cite{deUrreta:2013koa}. 
The operator
coefficients of the best fit are also compared with our fits and non-trivial
correlations among the coefficients that connect them with the details of the
microscopic quark-quark interaction are uncovered.   

The paper is organized as follows. In Sec.~\ref{sec:quarkmodel} we present the
physical states and define the mixing angles. In Sec.~\ref{sec:matrixelements}
we present the model interactions. In Sec.~\ref{sec:massmat} we present the
general parametrization of the mass matrix. In Sec.~\ref{sec:marel} we present
the mass-angle relations for the general case and for the more restrictive OGE,
OME model interactions. In Sec.~\ref{sec:fits} we discuss the fits and in
Sec.~\ref{sec:Nc} the connection with the $1/N_c$ operators. In
Sec.~\ref{sec:conclusions} we present our conclusions.  In App.~\ref{CCGLops}
we list the $1/N_c$ operators and in App.~\ref{oprels} we show the linear
dependences relevant for $N_c=3$. The detailed analytic expressions that
connect our results with the $1/N_c$ studies are presented in
App.~\ref{cbasis}.  In App.~\ref{angledep} we give the explicit expressions of
the coefficients as a function of the angles.  Useful expressions for the
general form of the mass matrix and its relation to the mixing angles are given
in App.~\ref{massmatdiag}.  In App.~\ref{IKV}  we cross check our general
expressions and show how they reduce in the simplest version of the QM, the IK
model. 

\section{Quark model and physical states}
\label{sec:quarkmodel}

We consider a non-relativistic quark model for baryons, described by 
the Hamiltonian $H_0$ for three quarks of constituent
mass $m$ with two-body harmonic interactions
\begin{eqnarray} 
H_0 = \frac{1}{2m} \sum_i p_i^2 + \frac{K}{2}
\sum_{i<j} r_{ij}^2 \,, 
\end{eqnarray} 
where $K$ is  a model parameter.  This Hamiltonian can be diagonalized exactly
and the center-of-mass (CoM) motion decouples when $H_0$ is expressed in terms
of the relative coordinates $\vec \rho = \frac{1}{\sqrt2}(\vec r_1
- \vec r_2)$ and  $\vec \lambda = \frac{1}{\sqrt6}(\vec r_1 + \vec r_2 - 2 \vec
  r_3)$.
The Hamiltonian takes the form 
of two independent oscillators
\begin{eqnarray} 
H_0 = \frac{p_\rho^2}{2 m} + \frac{p_\lambda^2}{2 m} +
\frac32 K \rho^2 + \frac32 K \lambda^2 \,,
\end{eqnarray} 
where $\vec p_\rho = -i \, \partial/\partial \vec \rho  , \ \vec p_\lambda = -i
\, \partial/\partial \vec \lambda $.
  
Here we will consider the first excited
states carrying one unit of angular momentum, which correspond to the lowest
energy negative parity baryons that belong to the mixed symmetric (MS) ${\bf
20_{MS}}$ multiplet of spin-isospin $SU(4)$.
The eigenstates $\Psi^{\rho,\lambda}_{Lm}$ of $H_0$ with
$L=1,m=1$ are 
\begin{eqnarray} \label{psilambda1}
\Psi^\rho_{11} &=& \rho_+
\frac{\alpha^4}{\pi^{3/2}}\exp\left(-\frac12 \alpha^2 (\rho^2 +
\lambda^2)\right) \ , \\ 
\Psi^\lambda_{11} &=& \lambda_+
\frac{\alpha^4}{\pi^{3/2}}\exp\left(-\frac12 \alpha^2 (\rho^2 +
\lambda^2)\right) \,, \label{psilambda2}
\end{eqnarray} 
where $\alpha = (3 K m)^{1/4}$, $\rho_+ = \rho_x + i \rho_y$, $\lambda_+ = \lambda_x + i
\lambda _y$ and the combination 
$\rho^2 + \lambda^2$ is invariant under permutations of the three quarks. 

The quark spin of the MS $I=1/2$ states takes the values  $S=1/2, 3/2$, which
combined with the orbital angular momentum $L=1$ gives the following $N$
states: two states with $J=1/2$ denoted $^2N_{1/2}, {^4N}_{1/2}$, two states $J=3/2$
denoted $^2N_{3/2}, {^4N}_{3/2}$, and one state with $J=5/2$ denoted $^4N_{5/2}$. In
addition, there are also spin-isospin MS $I=3/2$ states with $S=1/2$, which
result in  two $J=1/2,3/2$ states, denoted as $^2\Delta_{1/2}, {^2\Delta}_{3/2}$.

The MS spin-isospin states are coupled with the MS spatial states given in
Eqs.~(\ref{psilambda1},\ref{psilambda2}) to form completely symmetric states with the right quantum numbers,
antisymmetrized by the color singlet wavefunction, which factorizes from all our
calculations and can be omitted. 
Explicitly the MS quark model states $^{2 S+1}N_J$ and $^{2 S+1}\Delta_J$ are given by 
\begin{eqnarray}
|^2N_J;J_3 I_3 \rangle &=& \frac12 \sum_{m,S_3}
\left(
\begin{array}{cc|c}
1 & \frac12 & J \\
m & S_3 & J_3 
\end{array}
\right) 
\left[ 
\left(
\xi^\rho_{S_3} \varphi^\rho_{I_3} - \xi^\lambda_{S_3} \varphi^\lambda_{I_3}
\right) \Psi^\lambda_{1m} + 
\left(
\xi^\rho_{S_3} \varphi^\lambda_{I_3} - \xi^\lambda_{S_3} \varphi^\rho_{I_3}
\right) \Psi^\rho_{1m}
\right] , \\
|^4N_J;J_3 I_3 \rangle &=& \frac{1}{\sqrt2}  \sum_{m,S_3}
\left(
\begin{array}{cc|c}
1 & \frac32 & J \\
m & S_3 & J_3 
\end{array}
\right) 
\xi^{3/2}_{S_3}
\left(
\varphi^\rho_{I_3} \Psi^\rho_{1m} + \varphi^\lambda_{I_3} \Psi^\lambda_{1m}
\right)  \ , \\
|^2\Delta_J;J_3 I_3 \rangle &=& \frac{1}{\sqrt2} \ \varphi^{3/2}_{I_3}  \sum_{m,S_3}
\left(
\begin{array}{cc|c}
1 & \frac12 & J \\
m & S_3 & J_3 
\end{array}
\right) 
\left(
\xi^\rho_{S_3} \Psi^\rho_{1m} + \xi^\lambda_{S_3} \Psi^\lambda_{1m}
\right)
\ ,
\end{eqnarray}
where $\varphi$ and $\xi$ are isospin and spin states that are orthonormal. The
explicit spin states relevant for the calculation are
\begin{eqnarray} 
\xi^{3/2}_{3/2} &=&| \uparrow \uparrow \uparrow \rangle  \ , \\
\xi^\rho_{1/2} &=& \frac{1}{\sqrt{2}} 
\left( | \uparrow \downarrow \uparrow \rangle - | \downarrow \uparrow \uparrow \rangle \right) \ , \\ 
\xi^\lambda_{1/2} &=& - \frac{1}{\sqrt{6}} 
\left( | \uparrow \downarrow \uparrow \rangle 
     + | \downarrow \uparrow \uparrow \rangle 
     - 2  | \uparrow \uparrow \downarrow \rangle \right) \ , 
\end{eqnarray} 
where the other spin projections can be obtained by applying the lowering operator, and similarly 
for isospin. 

The physical states of angular momentum $J$ are a mixture of quark model states, as the quark spin $S$ is 
not a good quantum number.  
The mixing angles are defined as
\begin{eqnarray} 
N_{J=1/2}(1535) &=& \;\; \cos\theta_{1} \; ^2N_{1/2} + \sin\theta_{1} \; ^4N_{1/2} \,,
\nonumber  \\ 
N'_{J=1/2}(1650) &=& -\sin\theta_{1} \;  {^2}N_{1/2} + \cos\theta_{1} \; ^4N_{1/2} \,, 
\end{eqnarray} 
for
the $J=1/2$ nucleons, and 
\begin{eqnarray} 
N_{J=3/2}(1520) &=&  \;\; \cos\theta_{3} \; ^2N_{3/2}+ {\sin\theta_{3}} \;  {^4N}_{3/2}
\,, \nonumber \\
N'_{J=3/2}(1700) &=& -{\sin\theta_{3}} \; ^2N_{3/2} + {\cos\theta_{3}} \;  ^4N_{3/2} \,, 
\end{eqnarray} 
for the $J=3/2$ nucleons \footnote{Notice the change of notation respect to
Ref.~\cite{Carlson:1998vx} regarding the meaning of the prime on states.
Instead of $S=3/2$ quark model states, $N'_J$ are now physical states.}.
It is always possible to bring the mixing angles into the range $[-\pi/2 ,
\pi/2]$ by appropriate phase redefinitions of the physical states, and this is
the definition we will use throughout this paper.

\section{Model interactions and their spin-flavor structure }
\label{sec:matrixelements}

The spin-isospin symmetric Hamiltonian $H_0$ contributes an average mass $m_0$ to all members
of  an  $SU(4) \times O(3)$ multiplet. In order to describe the mass 
splittings of the negative parity baryons we will add to $H_0$ the spin-isospin dependent two-body
interaction terms $V_{ij}$
\begin{eqnarray}\label{Hamiltonian}
H = H_0 + \sum_{i<j} V_{ij} = H_0 + V_{ss} + V_{so} + V_{t} \, , 
\end{eqnarray}
which are labeled according to their transformation properties under orbital
angular momentum $\ell$: the $\ell=0$ spin-spin interaction $V_{ss}$, the
$\ell=1$ spin-orbit interaction $V_{so}$ and the $\ell=2$ quadrupole
interaction $V_t$.  We write these interaction terms in a generic form,
leaving their radial dependence unspecified, following
Refs.~\cite{Collins:1998ny,Pirjol:2007ed} as 
\begin{eqnarray}
\label{vss}
V_{ss} = V^0_{ss} + V^1_{ss}  &=& \sum_{i < j=1}^3 v_{ss}(r_{ij}) \vec s_i \cdot \vec s_j
\,, \\
\label{vso}
V_{so} = V^0_{so} + V^1_{so}  &=& \sum_{i < j=1}^3 v_{so}(r_{ij}) \Big[ 
(\vec r_{ij} \times \vec p_i ) \cdot \vec s_i - (\vec r_{ij} \times \vec p_j ) \cdot \vec
s_j \nonumber\\
& & \hspace{2.5cm}+ 2 (\vec r_{ij} \times \vec p_i ) \cdot \vec s_j - 2 
(\vec r_{ij} \times \vec p_j ) \cdot \vec s_i \Big]  \,, \\
\label{vt}
V_{t} = V^0_{t} + V^1_{t}  &=& \sum_{i < j=1}^3 v_{t}(r_{ij}) \Big[3(\hat r_{ij} \cdot
\vec s_i)(\hat r_{ij} \cdot \vec s_j)
 -  (\vec s_i \cdot \vec s_j) \Big]  \, , 
\end{eqnarray}
with $v_\varkappa(r_{ij}) = v^0_\varkappa(r_{ij}) + v^1_\varkappa(r_{ij}) \tau^a_i
\tau^a_j $,  where $\varkappa=ss,so,t$, labels the $v^{0,1}_\varkappa$ radial
functions that we take as unknown.  The superscript $0 (1)$ indicates the
absence (presence) of the isospin dependent $\tau_i \cdot \tau_j$ interactions.
Setting $v^{1}_\varkappa=0$ ($v^{0}_\varkappa=0$) we obtain the isospin
independent (dependent) interactions that we label as OGE (OME), respectively,
as they can be typically obtained starting from one-gluon exchange 
\cite{De Rujula:1975ge} or one-meson exchange \cite{Glozman:1995fu}
model interactions. We label the general quark model as defined by Eqs.~(\ref{Hamiltonian})-(\ref{vt})
as the OGE + OME quark model.

As is well known \cite{Isgur:1978xj}, although we start from two-body
interactions, it is remarkable that after removing the CoM motion 
the spin-orbit potential develops a three-body interaction, as can be seen from
its $(ij)=(12)$ component
\begin{eqnarray}
(V_{so})_{12} &=& 3 \; v_{so}( \sqrt{2} \rho ) 
\Big[ 
(\vec \rho \times \vec p_\rho ) \cdot (\vec s_1 + \vec s_2) 
- \frac{1}{3 \sqrt{3}}  (\vec \rho \times \vec p_\lambda ) \cdot (\vec s_1 -  \vec s_2) 
\Big] \\
&\equiv& (V_{so-2B})_{12}+ (V_{so-3B})_{12} \ , \nonumber
\end{eqnarray}
where the two-body component $V_{so-2B}$ and the three-body component  $V_{so-3B}$ are defined by the 
first and second term,  respectively. 

\section{Mass matrix parametrization and operator matching }
\label{sec:massmat}

Here we present the general structure we obtain for the mass matrix of the
$[{\bf 20_{MS}}, 1^-]$ non-strange excited baryons starting from the generic
quark model defined by Eqs.~(\ref{Hamiltonian})-(\ref{vt}).  We also discuss
the matching to the effective spin-flavor operators that appear in the studies
of excited baryons using the $1/N_c$ expansion, as defined in
Ref.~\cite{Carlson:1998vx}.  They will be labeled as ``CCGL operators'' and can
be found listed again in Appendix~\ref{CCGLops} for the convenience of the
reader. 

In the quark model, the matrix elements are obtained by explicit computation
using the harmonic oscillator basis of eigenstates of $H_0$ with the help of
(see e.g.  Ref.~\cite{Edmonds}):
\begin{eqnarray}
\langle \gamma' j_1' j_2' J M | \vec T(k) \cdot \vec U(k) | \gamma j_1 j_2 J M \rangle 
&=&
(-)^{j_1+j_2'+J} 
\left\{
\begin{array}{ccc}
J & j_2' & j_1' \\
k & j_1  & j_2 
\end{array}
\right\}
\sum_{\gamma''} 
\langle \gamma'  j_1' || \vec T(k) || \gamma'' j_1 \rangle 
\langle \gamma'' j_2' || \vec U(k) || \gamma   j_2 \rangle 
\ , \nonumber
\end{eqnarray}
where $\langle \dots || \dots || \dots \rangle $ stands for the reduced matrix
elements, the expression in brackets is a 6j-symbol and the dot in $ \vec T(k)
\cdot \vec U(k)$ means that all spatial indices of the tensor operators $\vec
T(k)$ and $\vec U(k)$ of rank $k$ are contracted.  The reduced matrix elements of the
spatial operator can be expressed in terms of radial integrals of the unknown
functions $v^{0,1}_\varkappa$.

Our results for the matrix elements of the OGE + OME model
$H = H_0 + V_{ss} + V_{so} + V_{t} $ in the quark model states basis are given in Table~\ref{table1}, from
where we can also read off the matrix elements of the more restricted OGE and OME
models.  The off-diagonal matrix elements are denoted as $^2N_J - {^4N_J}$. 

\begin{table}[h]
$
\begin{array}{crrrrrrr}
\hline  
\hline  
              &  H_0        & V_{ss}^0 & V_{ss}^1 & V_{so-2B}^{0,1}    & V_{so-3B}^0  & V_{so-3B}^1       & V_t^{0,1}     \\  
\hline  
^2N_{1/2}      & m_0         &  S^0     & S^1     & 2 \, P_{2B}^{0,1}  &  0           &    -4 \, P_{3B}^1    & 0     \\
^4N_{1/2}    & m_0           & -S^0     & S'^1     & 5 \, P_{2B}^{0,1}  &  0           &  0 & 5 \, D^{0,1}   \\
\ \  ^2N_{1/2} - {^4N}_{1/2} \ \  & 0 & 0\;\;    & 0\;\;\;  & P_{2B}^{0,1}      &  P_{3B}^0    & {P}_{3B}^1      & 5 \, D^{0,1} \\
^2N_{3/2}              & m_0 &  S^0     & S^1     & - \, P_{2B}^{0,1}  &  0           &  2 \, P_{3B}^1     & 0                 \\
^4N_{3/2}              & m_0 & -S^0     & S'^1     & 2 \, P_{2B}^{0,1}  &  0           &  0 & -4 \, D^{0,1}              \\
^2N_{3/2} - {^4N}_{3/2} & 0& \ \ \ \ \ \   0\;\; & 0\;\;\; & \ \ \  \sqrt{\frac52}\,P_{2B}^{0,1} 
& \ \sqrt{\frac52}\,P_{3B}^{0} & \ \sqrt{\frac52}\,{P}_{3B}^1 & \ \ \ - \sqrt{\frac52}\,D^{0,1}  \\
^4N_{5/2}              & m_0 & -S^0     & S'^1     & -3 \, P_{2B}^{0,1} &  0           &  0 &   D^{0,1}                 \\
^2\Delta_{1/2}        & m_0  &  S'^0    & S'^1     & 0                  & -2\, P_{3B}^0 &  2 \, {P}_{3B}^1  & 0  \\ 
^2\Delta_{3/2}        & m_0  &  S'^0    & S'^1     & 0                  & P_{3B}^0
&   - \, {P}_{3B}^1   & 0               \\ 
\hline  
\hline  
\end{array} 
$
\caption{ Matrix elements of the spin-spin $V_{ss}$, two-body spin-orbit
$V_{so-2B}$, three-body spin-orbit $V_{so-3B}$, and tensor $V_{t}$ component of
the interaction. The 0 and 1 superscripts correspond to the OGE and OME case,
respectively.  }
\label{table1} 
\end{table}

\begin{table}[h]
$
\begin{array}{crrr} 
                        &  O_{\ell=0} & O_{\ell=1}        & O_{\ell=2} \\  
\hline  
\hline  
^2N_{1/2}               &      S_1    & -2 \,  P_1           &  0                   \\
^4N_{1/2}               &      S_2    & 5 \, P_2             &  5  \, D_1           \\
^2N_{1/2} - {^4N}_{1/2} &      0      &   P_3                & \; -5 \, D_2          \\
^2N_{3/2}               &      S_1    &   P_1            &  0         \\
^4N_{3/2}               &      S_2    &  2 \, P_2            &  -4  \, D_1          \\
^2N_{3/2} - {^4N}_{3/2} & \ 0           & \ \ \sqrt{\frac52}\,P_3   & \ \ \sqrt{\frac52}\,D_2  \\
^4N_{5/2}               &     S_2     &  -3 \, P_2           &  D_1                 \\
^2\Delta_{1/2}          &     S_3     &  -2 \, P_4           &  0                   \\
^2\Delta_{3/2}          &     S_3     &  P_4                 &  0 \\
\hline  
\hline  
\end{array} 
$
\caption{General structure of the $\ell=0,1$ and $2$ mass matrix expressed in terms of 
the CCGL operators that appear in the $1/N_c$ expansion.}
\label{table1b} 
\end{table}

The matrix elements shown in Table~\ref{table1} involve $m_0$ and ten constants
that are expressed in terms of the radial integrals $I_2, I_4, U$ and $J_4$ of
the unknown functions $v^{0,1}_{\varkappa}$ as
\begin{equation}
 \begin{aligned}[c]
  S^0     =&-\frac{3}{2}\frac{\alpha^3}{\sqrt{\pi}}\left( I^0_2+\frac{2}{3}\alpha^2
I^0_4\right) \ , \\
  S'^0    =& \frac{3}{2}\frac{\alpha^3}{\sqrt{\pi}}\left( I^0_2-2\alpha^2 I^0_4\right) \ , \\
  S^1     =&\frac{3}{8}\frac{\alpha^3}{\sqrt{\pi}} \left(5 I^1_2-2\alpha^2 I^1_4\right) \ ,   \\
  S'^1    =&\frac{3}{8}\frac{\alpha^3}{\sqrt{\pi}} \left(  I^1_2-2\alpha^2 I^1_4\right) \ , \\
 \end{aligned}
 \qquad
 \qquad
 \begin{aligned}[c]
  P^0_{2B}=&-4 \frac{\alpha^5}{\sqrt{\pi}} U^{0} \ ,  \\
  P^0_{3B}=& \frac{8}{27}\frac{\alpha^5}{\sqrt{\pi}} U^{0} \ ,  \\
  P^1_{2B}=& 3 \frac{\alpha^5}{\sqrt{\pi}} U^{1} \ , \\
  P^1_{3B}=&-\frac{2}{27}\frac{\alpha^5}{\sqrt{\pi}} U^{1} \ , \\
 \end{aligned}
 \qquad
 \qquad
 \begin{aligned}[c]
  D^0     =&-\frac{2}{5}\frac{\alpha^5}{\sqrt{\pi}} J^0_4 \ , \\
  D^1     =& \frac{3}{10}\frac{\alpha^5}{\sqrt{\pi}} J^1_4 \ ,  
 \end{aligned}
\label{SPDIUJ}
\end{equation}
where the radial integrals are
\begin{equation}
 \begin{aligned}[c]
   I^{0,1}_2=\int_0^\infty \rho^2 v^{0,1}_{ss}(\sqrt{2}{\rho}) e^{-\alpha^2\rho^2}d\rho \ , \\
   I^{0,1}_4=\int_0^\infty \rho^4 v^{0,1}_{ss}(\sqrt{2}{\rho})e^{-\alpha^2\rho^2}d\rho \ ,  \\
   \end{aligned}
 \qquad
 \qquad
 \begin{aligned}[c]
   J^{0,1}_4=\int_0^\infty \rho^4 v^{0,1}_t(\sqrt{2}{\bf \rho})e^{-\alpha^2\rho^2}d\rho \ , \\
   U^{0,1}=\int_0^\infty\rho^4 v_{so}^{0,1}(\sqrt{2}{\rho})e^{-\alpha^2\rho^2}d\rho \ ,  
 \end{aligned}
\label{integrals}
\end{equation}
and depend on the particular shapes and relative strengths of the spin-flavor
interactions of a given model. These interactions will give the mass splittings
within a spin-flavor multiplet.

The physical masses and mixing angles are obtained diagonalizing the mass
matrix, whose general form in the quark model basis of $S=1/2, 3/2$ quark spin
states  and its relation to the mixing angles can be found in
Appendix~\ref{massmatdiag}.  

Notice that the tensor interaction and the two-body part of the spin-orbit
interaction of the OGE and OME models are proportional to each other (i.e.
$V^1_{so-2B} \sim V^0_{so-2B}$ and $V^1_t \sim V^0_t$).  This allows us to list
them in just one column in Table~\ref{table1}, as they only differ in the
numerical values of $D^0$, $D^1$ and $P_{2B}^0$, $P_{2B}^1$, respectively.

Alternatively, the mass matrix can also be written as a linear combination of
the 18 spin-isospin CCGL operators~\cite{Carlson:1998vx}, 
see Appendix~\ref{CCGLops}.
These operators were
originally constructed to perform a $1/N_c$ expansion analysis, but they are
also useful at fixed $N_c=3$, since they provide an overcomplete basis for the
mass matrix of the physical baryon states, that allows to factor out explicitly
the radial dependence of the states and interactions, as will be seen
explicitly as a result of the matching to the quark model results.  Grouping
the effective $\ell=0,1,2$ operators together we obtain the mass operator as  
\begin{eqnarray}
M &=& \sum_{i=1}^{18} c_i \, O_i = O_{\ell=0} +  O_{\ell=1} + O_{\ell=2} \ , 
\end{eqnarray}
where
\begin{eqnarray}
O_{\ell=0} &=& \sum_{i=1,6,7,11,16} c_i \, O_i \ ,  \\
O_{\ell=1} &=& \sum_{i=2,4,5,9,10,13,14,15} c_i \, O_i \ , \\
O_{\ell=2} &=& \sum_{i=3,8,12,17,18} c_i \, O_i \ , 
\end{eqnarray}
and the operator coefficients $c_i$ are numbers that are usually determined by 
fitting to data. 
We observe that the matrix elements of these operators, whose explicit
expressions can be found in Ref.~\cite{Carlson:1998vx}, have the general
structure shown in Table~\ref{table1b}. For $\ell=0$ they can be
parametrized by three parameters $S_{1,2,3}$, for $\ell=1$ they can be
parametrized by four parameters $P_{1,2,3,4}$ and for $\ell=2$ they can be
parametrized by the two parameters $D_{1,2}$.

In Appendix~\ref{oprels} we give the linear relations satisfied by the
operators on the nine-dimensional subspace of physical masses and mixing
angles at $N_c=3$, showing explicitly that  there are only three, four and two
independent operators (out of five, eight and five) for $\ell=0,1,2$,
respectively.

The relations between the coefficients $c_i$ and the parameters $S,P,D$ are
given in Appendix~\ref{cbasis}.  Comparing Table~\ref{table1b} and
Table~\ref{table1} we find 
\begin{eqnarray}
\label{eqs1}
S_1 &=&  m_0 + S^0 + S^1  \ , \\
S_2 &=&  m_0 - S^0 + S'^1  \ , \\
S_3 &=&  m_0 + S'^0 + S'^1 \ , \\
P_1 &=&  - P_{2 B}^{0} - P_{2 B}^{1} + 2 \, P_{3 B}^{1} \ ,  \\
P_2 &=&   P_{2 B}^{0} + P_{2 B}^{1} \ , \\
P_3 &=&   P_{2 B}^{0} + P_{2 B}^{1} + P_{3 B}^{0} + P_{3 B}^{1} \ , \\
P_4 &=&   P_{3 B}^{0} - P_{3 B}^{1} \ , \\
D_1 &=&   D^{0} + D^{1} \ , \\
D_2 &=&   -D^{0} -  D^{1} \ . 
\label{eqd2}
\end{eqnarray}

This, together with Eqs.~(\ref{eqC1})-(\ref{eqC9}) and
Eqs.~(\ref{SPDIUJ})-(\ref{integrals}),  establishes the complete analytic
matching of the generic quark model defined by
Eqs.~(\ref{Hamiltonian})-(\ref{vt}) to the spin-isospin effective operator
expansion used in the $1/N_c$ studies of excited baryons.  
A simple version of the quark model like the IK model provides a useful check
of these expressions, as summarized in Appendix~\ref{IKV}.

For the OGE+OME model and the more restrictive OGE, OME model interactions
there are constraints on  these nine parameters leading to the parameter-free
relations between the mixing angles and physical masses that will be discussed
in the next Section.

\section{Mass-angle relations}
\label{sec:marel}

The general parametrization of the mass matrix shown on Table~\ref{table1b}
involves nine parameters: three for the scalar part, four for the vector part
and two for the tensor part of the interaction.  These nine parameters can be
solved in terms of the physical masses and mixing angles, which will allow us to
study their variation on the mixing angles in Sec.~\ref{sec:fits}.

However, for the most general OGE+OME quark model the eleven constants that
give all the matrix elements in Table~\ref{table1} only appear as seven
independent combinations,  as can be seen explicitly from
Eqs.~(\ref{eqs1})-(\ref{eqd2}) and the constraints given by
Eq.~(\ref{relP-OGEOME}) and Eq.~(\ref{relD-OGEOME}) that will be discussed
next.  This implies that there must be two parameter-free relations among the
masses and mixing angles. 

The spin-orbit matrix elements of the OGE+OME model satisfy 
\begin{eqnarray}\label{relP-OGEOME}
P_1+2 P_2 - P_3 + P_4 &=& 0 \ ,  
\end{eqnarray}
implying our first mass-angle relation R1
\begin{eqnarray}\label{rel1}
\mbox{R1:} \qquad && \frac12 (N_{1/2}-N'_{1/2}) (3\cos 2\theta_{1} + \sin 2\theta_{1}) 
+ (N_{3/2}-N'_{3/2}) \left(-\frac35 \cos 2\theta_{3} + 
\sqrt{\frac52} \sin 2\theta_{3} \right) \nonumber \\
&& = - \frac12 (N_{1/2}+N'_{1/2}) + \frac75 (N_{3/2}+N'_{3/2})  - \frac95 N_{5/2} -
2\Delta_{1/2} + 2\Delta_{3/2} \,.
\end{eqnarray}
This relation was found for the first time in Ref.~\cite{Pirjol:2008gd},  see Eq.~(5) in this reference,
where it was
shown to hold for the most general quark model with two-body interactions.
It was also pointed out that the same relation is obtained in the $1/N_c$ expansion by
keeping all CCGL operators up to order $1/N_c$, using $N_c=3$ to evaluate their matrix elements.
The angle correlation it implies was already discussed in Fig.~5 of 
Ref.~\cite{Pirjol:2003ye}. 

\begin{figure}[tb]
\vskip 1 true cm
\includegraphics[width=7.5cm,keepaspectratio,angle=0,clip]{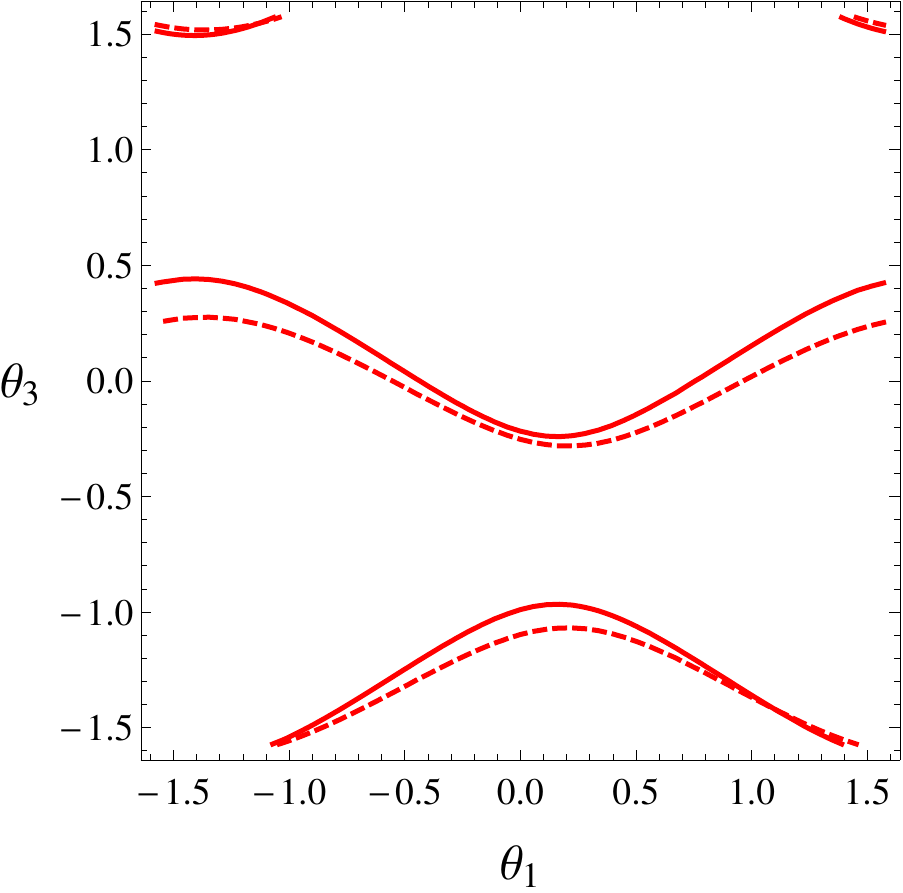}
	\caption{The angle correlation implied by mass relation R1, Eq.~(\ref{rel1}), is 
shown as a full line. The dashed line corresponds to 
R1', Eq.~(\ref{rel1p}), which differs from R1 by $1/N_c^2$ corrections.} 
\label{fig-urreta}
\end{figure}

More recently R1 was found again in the form
\begin{eqnarray}\label{rel1p}
\mbox{R1':} \qquad && (N_{1/2}-N'_{1/2}) (13\cos 2\theta_{1} + 4 \sqrt{2} \sin 2\theta_{1}) 
- 4 (N_{3/2}-N'_{3/2}) \left(\cos 2\theta_{3} - 
2 \sqrt{5} \sin 2\theta_{3} \right) \nonumber \\
&& = - 3 (N_{1/2}+N'_{1/2}) + 12  (N_{3/2}+N'_{3/2})  - 18  N_{5/2} -
24 \Delta_{1/2} + 24 \Delta_{3/2} \,,
\end{eqnarray}
in Refs.~\cite{deUrreta:2013koa,Fernando:2014dna} by expanding the matrix elements and 
dropping $1/N_c^2$ corrections in a $1/N_c$ analysis.
Both relations R1 and R1' imply correlations among the mixing angles that are
plotted in Fig.~\ref{fig-urreta} using the experimental masses of
Table~\ref{tab:exp}. Although the analytic expressions of R1 and R1' look very
different, numerically the angle correlation that follows from them at $N_c=3$
is very similar, a manifestation of the smallness of higher order $1/N_c^2$
corrections.

The second constraint in the OGE+OME case is: 
\begin{eqnarray}\label{relD-OGEOME}
D_1+D_2 &=& 0  \ , 
\end{eqnarray}
leading to the second mass-angle relation
\begin{eqnarray}\label{rel2}
\mbox{R2:} \qquad &&   5 (N_{1/2}-N'_{1/2}) (\cos 2\theta_{1} + 2 \sin 2\theta_{1}) 
  -  4 (N_{3/2}-N'_{3/2}) \left( 2 \cos 2\theta_{3} + \sqrt{\frac52} \sin 2\theta_{3}
     \right)  \nonumber \\
&& = 5 (N_{1/2}+N'_{1/2})  - 8 (N_{3/2}+N'_{3/2})  + 6 N_{5/2} \;. 
\end{eqnarray}

\begin{figure}[tb]
\vskip 1 true cm
\includegraphics[width=7cm,keepaspectratio,angle=0,clip]{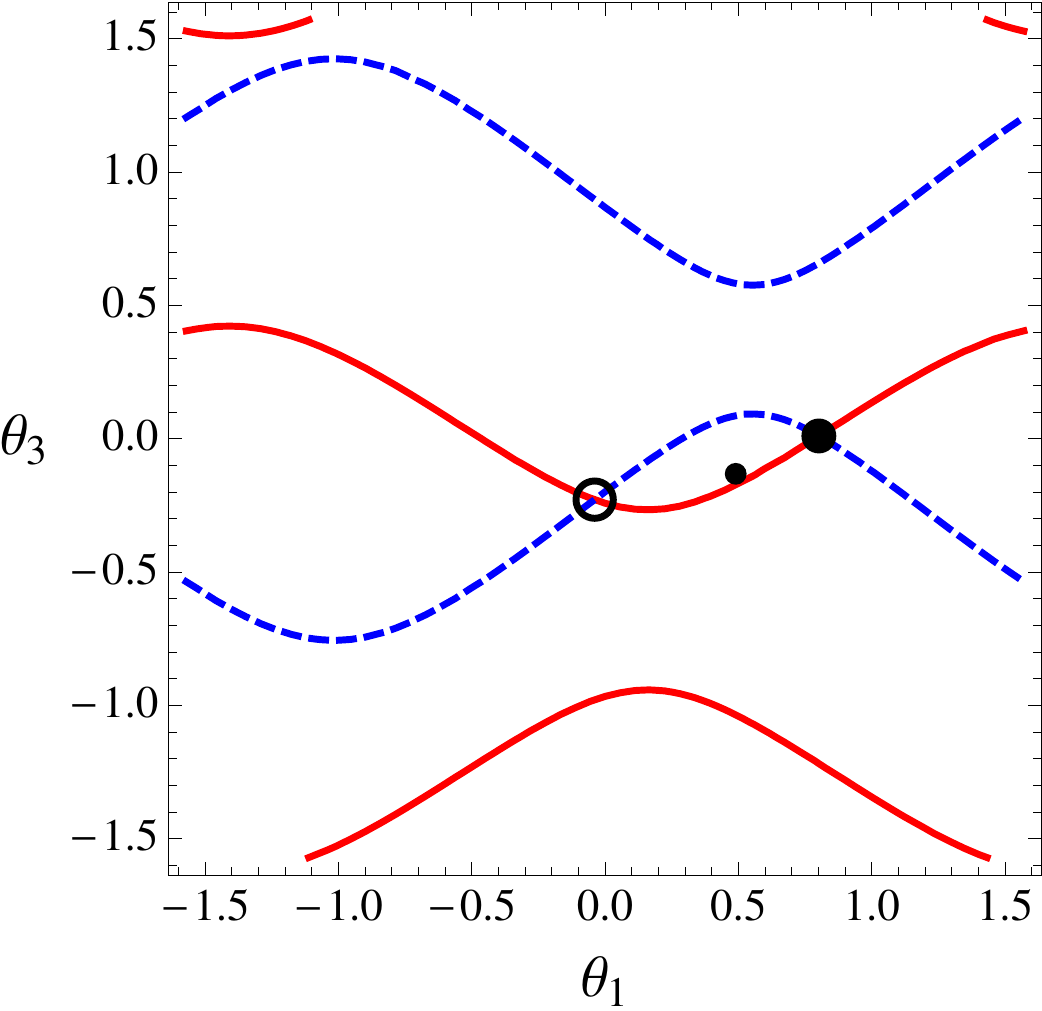}
\qquad
\qquad
\includegraphics[width=7cm,keepaspectratio,angle=0,clip]{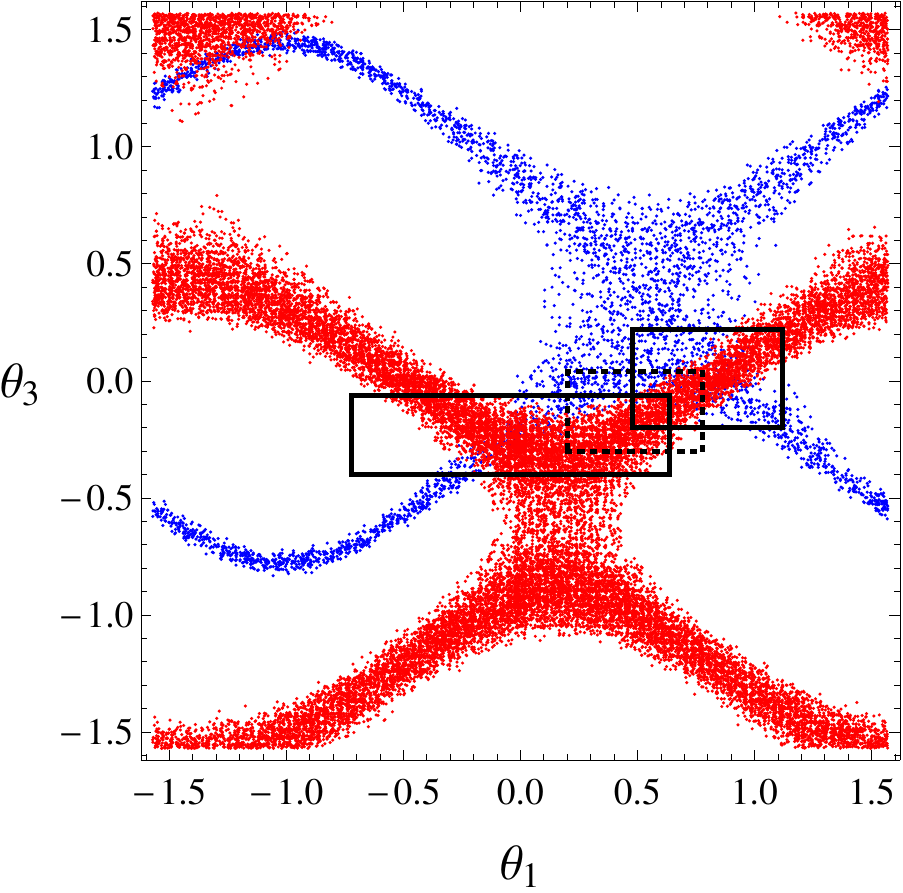}
	\caption{
Angle correlations from the relations R1 and R2, given by Eq.~(\ref{rel1}) and
Eq.~(\ref{rel2}).  Left panel: The curves giving the $\theta_1 - \theta_3 $ angle correlations
intersect at sol--A = $(0.80,0.01)$,
black dot,  and sol--B = $(-0.04, -0.23)$, circle, obtained using the central
values of the masses. The smaller point in between corresponds to the best fit
of the angles of  Ref.~\cite{deUrreta:2013koa}. 
Right panel:  scatter plot obtained including the experimental errors on the
masses. The rectangles correspond to sol--A, sol--B and the best fit
(dashed) with errors as in Table~\ref{table4}.
} 
\label{fig-r1r2}
\end{figure}

The two OGE+OME relations R1 and R2 are plotted  as the full and dashed curves
on the left panel of Fig.~\ref{fig-r1r2} using the central values for the
masses as given in Table~\ref{tab:exp}.  Each relation gives different
two-valued correlations between the two mixing angles $(\theta_1,\theta_3)$.
Their intersections give two possible solutions for the mixing angles, which we
label as sol--A= $(0.80 \pm 0.32,0.01 \pm 0.21)$, shown as a black dot,  for
the one with the larger value of $\theta_1$ and sol--B= $(-0.04 \pm 0.68,
-0.23 \pm 0.17)$, shown as a circle,  for the one with the smaller value of
$\theta_1$. The errors were obtained by propagating the errors of the
experimental masses as given in Table~\ref{tab:exp}.  The best values for the
mixing angles obtained in Ref.~\cite{deUrreta:2013koa} from a global fit to the
masses and decays are $ (0.49 \pm 0.29, -0.13 \pm 0.17)$, also shown in Fig.~\ref{fig-r1r2} 
as the smaller black dot.

\begin{table}
\begin{tabular}{cccccccc} 
\hline  
\hline  
      & \ $N_{1/2}(1535)$ \ & \ $N'_{1/2}(1650)$     \ & \ $N_{3/2}(1520)$      \ & \ $N'_{3/2}(1700)$ \ &
        \ $N_{5/2}(1675)$ \ & \ $\Delta_{1/2}(1620)$ \ & \ $\Delta_{3/2}(1700)$ \ \\ 
\hline
PDG(2014) & $1535\pm 10$ & $1658\pm 13$ & $1515\pm 5$ & $1700\pm 50$ & $1675 \pm 5$ & $1630\pm 30$ & $1710\pm 40$ \\ 
OGE       & $1533\pm37$ & $1659\pm43$ & $1516\pm36$ & $1717\pm19$ & $1675\pm16$ & $1627\pm39$ & $1716\pm30$  \\
OME       & $1535\pm26$ & $1659\pm23$ & $1515\pm19$ & $1693\pm17$ & $1675\pm19$ & $1637\pm18$ & $1683\pm15$  \\
\hline
\hline 
\end{tabular} 
\caption{The experimental values shown on the first line are taken from
Ref.~\cite{Agashe:2014kda}. The mass spectrum as obtained from the
fits to the OGE and OME interactions, as
discussed in Sec.~\ref{sec:fits}, is shown in the last two lines. Masses are given in MeV. }
\label{tab:exp} 
\end{table}

When the uncertainties in the masses are taken into account the curves on the angles plane
are expanded into bands, that are estimated in the scatter plot shown on the right panel. 
The rectangles correspond to sol--A, sol--B and the best fit angles with the error bars as in 
Table~\ref{table4}. In the upper left corner a third solution sol--C $\approx$  $(-1.14, 1.50 )$ 
is possible within the present experimental errors. It is 
similar to the $(-1.14, 1.40)$ solution $1'$ of the large $N_c$ analysis of Ref.~\cite{Pirjol:2003ye}, 
yet this solution is excluded by the 
best fit of Ref.~\cite{deUrreta:2013koa} that also includes information on the decay properties of 
the baryons. We therefore do not consider sol--C any further here.

\begin{figure}[tb]
\vskip 1 true cm
\includegraphics[width=7cm,keepaspectratio,angle=0,clip]{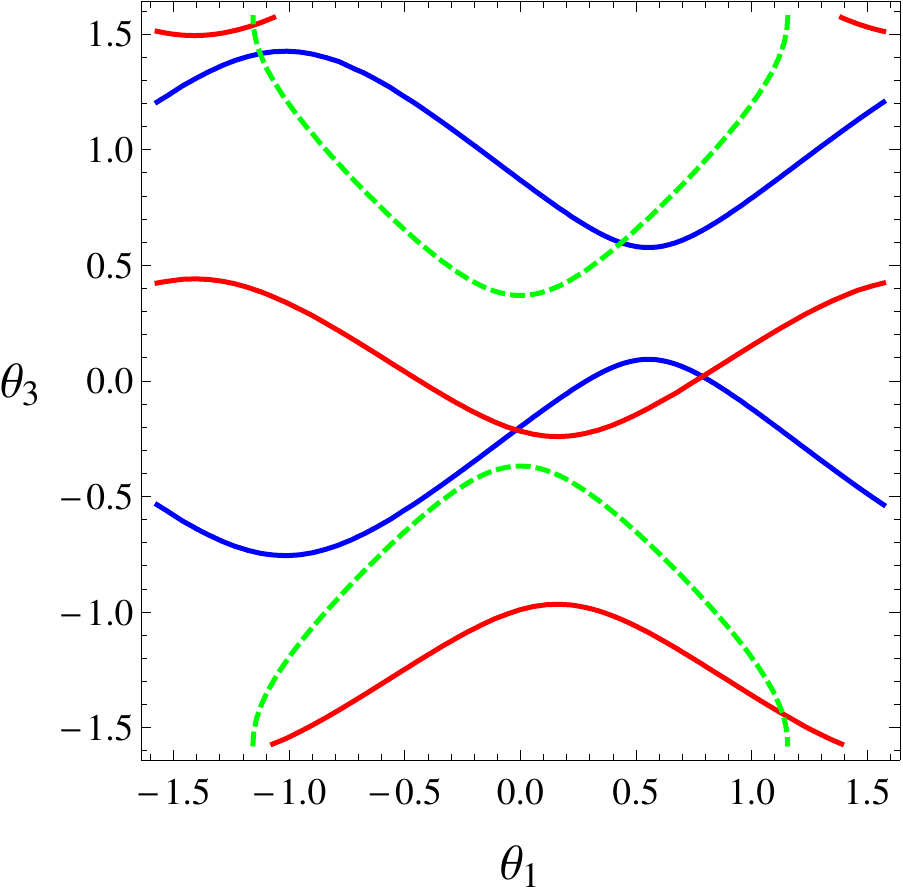}
\qquad
\qquad
\includegraphics[width=7cm,keepaspectratio,angle=0,clip]{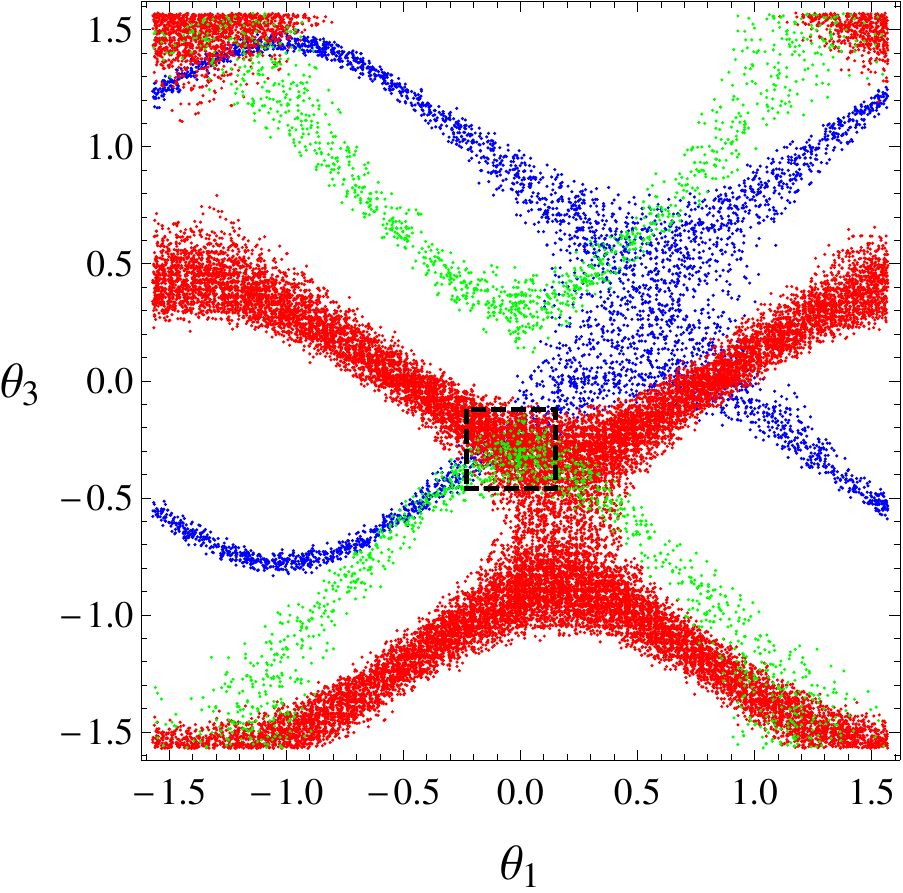}
	\caption{ 
(color online) OGE mass-angle relation R3,  Eq.~(\ref{rel4}) (dashed green) and
general correlations R1, Eq.~(\ref{rel1}) (red) and R2, Eq.~(\ref{rel2}) (blue)
Left: Using the central values of the masses. Right: Scatter plot obtained
including the experimental errors on the masses and OGE angles with errors (dashed rectangle) as
in Table~\ref{table4}.  } 
\label{fig-r1r2r3s1}
\end{figure}

If we restrict the interactions to the OGE case we have one additional relation among
matrix elements, namely
\begin{eqnarray}
P_1+ P_2 &=& 0 \ ,    \label{relP-OGE}
\end{eqnarray}
leading to 
\begin{eqnarray}\label{rel4}
\mbox{R3:} \qquad &&   -25 (N_{1/2}-N'_{1/2}) \cos 2\theta_{1}
     +16 (N_{3/2}-N'_{3/2}) \cos 2\theta_{3}  \nonumber \\
&& =  15 (N_{1/2}+N'_{1/2})  - 24 (N_{3/2}+N'_{3/2})  + 18  N_{5/2} \;.
\end{eqnarray}
The relations R1, R2 and R3 that hold in the OGE case  are shown in Fig.~\ref{fig-r1r2r3s1}
next to the corresponding scatter plot. Taken together they exclude sol--A.

\begin{figure}[tb]
\vskip 1 true cm
\includegraphics[width=7cm,keepaspectratio,angle=0,clip]{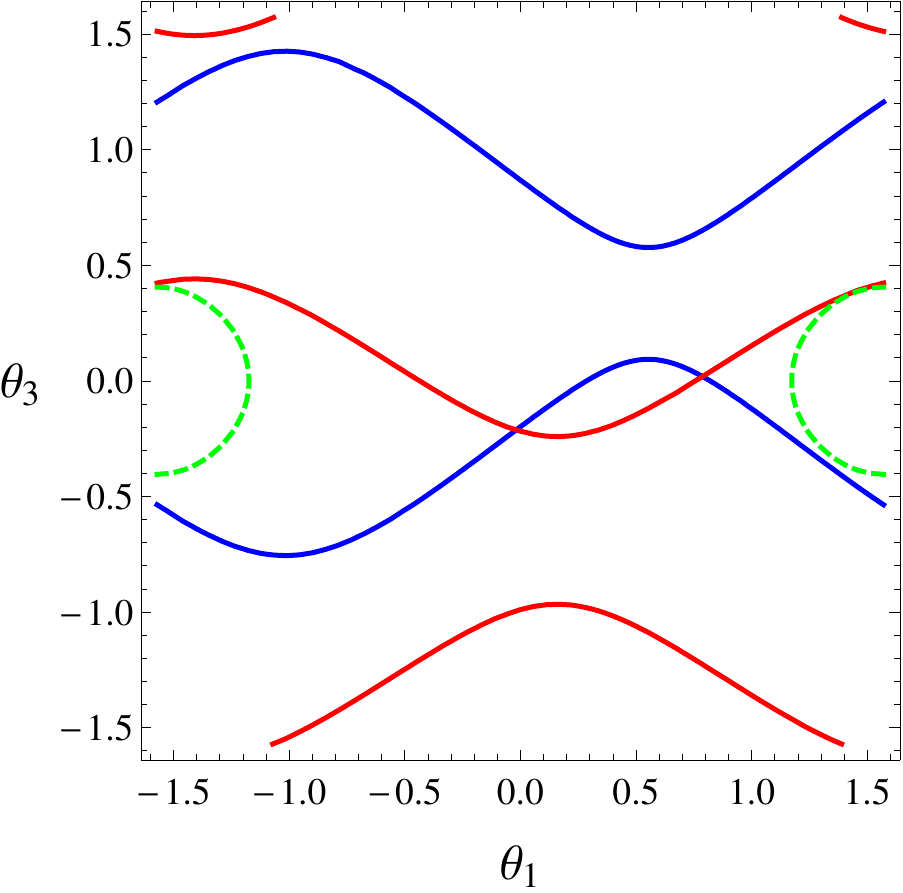}
\qquad
\qquad
\includegraphics[width=7cm,keepaspectratio,angle=0,clip]{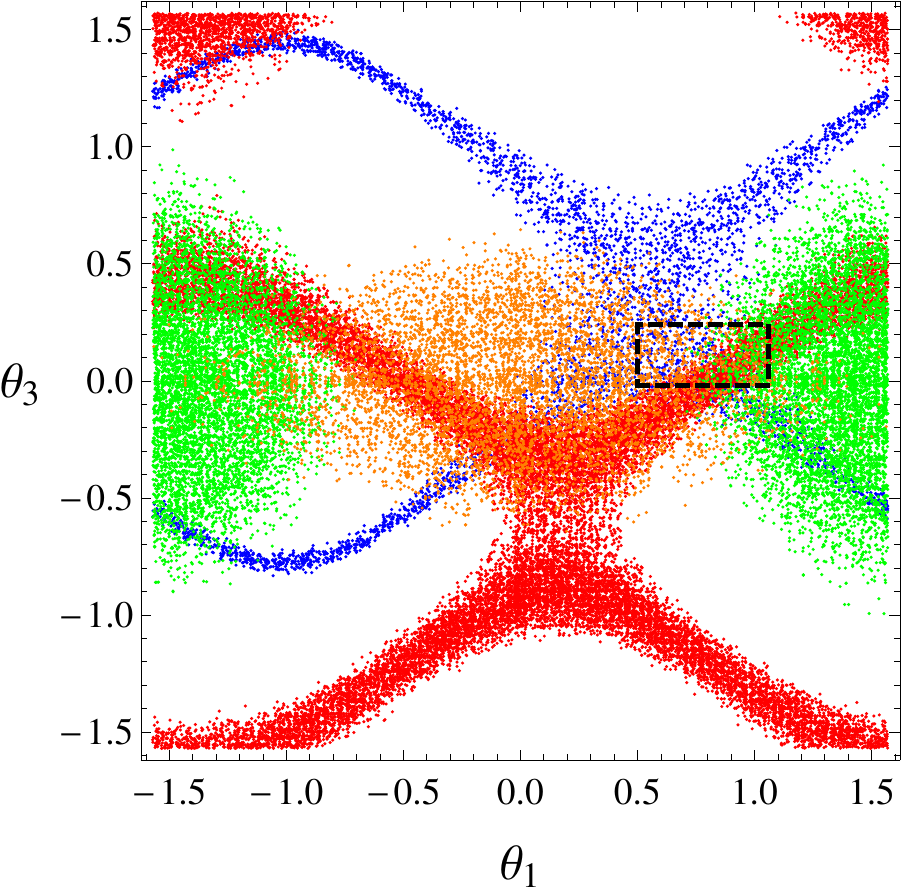}
	\caption{(color online) OME mass relations R4, Eq.~(\ref{rel5}) (orange) and R5, Eq.~(\ref{rel6})
(green),  and general correlations R1, Eq.~(\ref{rel1}) (red) and R2, Eq.~(\ref{rel2})
(blue). Left: Using the central values of the masses R4 is not
satisfied. Right: Scatter plot obtained including the experimental errors on
the masses and OME angles with errors (dashed rectangle) as
in Table~\ref{table4}.} 
\label{fig-r1r2r4}
\end{figure}

In the OME case, in addition to Eq.~(\ref{relP-OGEOME}) and Eq.~(\ref{relD-OGEOME}), we have 
two other constraints from the scalar and vector part of the interaction
\begin{eqnarray}
S_2-S_3  &=& 0        \ ,    \label{relS-OME} \\
P_1+ P_2 + 2 P_4 &=& 0  \ ,  \label{relP-OME}
\end{eqnarray}
giving the mass-angle relation
\begin{eqnarray}\label{rel5}
\mbox{R4:} \qquad &&    -  (N_{1/2}-N'_{1/2}) \cos 2\theta_{1}
     - 2(N_{3/2}-N'_{3/2}) \cos 2\theta_{3}  \nonumber \\
&& =  - (N_{1/2}+N'_{1/2})  - 2 (N_{3/2}+N'_{3/2})  - 6  N_{5/2} + 4\Delta_{1/2} +
8\Delta_{3/2} \ , 
\end{eqnarray}
and 
\begin{eqnarray}\label{rel6}
\mbox{R5:} \qquad &&   -25 (N_{1/2}-N'_{1/2}) \cos 2\theta_{1}
     +16 (N_{3/2}-N'_{3/2}) \cos 2\theta_{3}  \nonumber \\
&& =  15 (N_{1/2}+N'_{1/2})  - 24 (N_{3/2}+N'_{3/2})  + 18  N_{5/2} + 80 \Delta_{1/2} - 80
\Delta_{3/2} \; ,  
\end{eqnarray}
respectively.
In Fig.~\ref{fig-r1r2r4} we show R1, R2, R4 and R5 together with the
corresponding scatter plot.
It is important to stress that R4 is not only predicted in the pure
OME case, but also in the general OGE+OME interaction when the OGE part of the
spin-spin interaction is of zero range.  In that case $I^{0,1}_4=0$ gives
$S^0=-S'^0$, which implies $S_2=S_3$ and Eq.~(\ref{relS-OME}) holds for a generic OGE+OME
interaction as well. 
The orange area in the scatter plot of
Fig.~\ref{fig-r1r2r4} corresponds to the region were R4 is satisfied within
experimental errors and overlaps with the sol--A, sol--B and best fit
rectangles shown on  Fig.~\ref{fig-r1r2}, all compatible with a spin-spin interaction
with a flavor independent component of short range.

Finally, in order to quantify to what extent the relations R1...5 are
satisfied, we define their accuracy (in \%) as $ 100 \times
|LHS-RHS|/(LHS+RHS)$ from the left-hand-side ($LHS$) and right-hand-side ($RHS$) of
the equation $LHS=RHS$, corresponding to each of R1...5 written so that all terms in their 
$LHS,RHS$ are positive.
Taking typical values for the masses and splittings a good agreement is
signaled by an accuracy value of less than 1 \%.  Table~\ref{table7} shows that
R1, R2 are always satisfied, the OGE relation R3 is favored by sol--B and
sol--C, while the OME relations (R4 and R5) are favored by sol--A.  R4 is well
satisfied for both sol--A and sol--B, consistent with what was seen in the
scatter plot.  Again, this is compatible with a short range for the flavor independent
spin-spin component of the interaction.

\begin{table}
\begin{tabular}{ccccccc} 
\hline  
\hline 
& $(\theta_1,\theta_3)$ & R1 & R2 & R3 & R4 & R5 \\ 
\hline
sol--A & $(0.80,0.01)$  & 0.00    & 0.00    & 1.89  & 0.35 & 0.52 \\
sol--B & \ \ \  $(-0.04, -0.23)$ \ \ \ & 0.00    & 0.00    & 0.17 & 0.15  & 1.08 \\ 
sol--C & $(-1.14, 1.50 )$ & 0.15     & 0.23 & 0.01 & 2.01  & 1.22   \\
\hline 
\hline 
\end{tabular} 
\caption{Accuracy (in \% ) of the mass relations as defined in the text.}
\label{table7} 
\end{table}

\section{Mixing angles and best fits}
\label{sec:fits}

\begin{table} [h]
\begin{tabular}{ccccccccccc}
\hline
\hline
         & $S_0$ & $\Delta S_{12}$ & $\Delta S_{3}$ & $P_1$ & $P_2$ & $P_3$ & $P_4$ & $D_1$ & $D_2$   & $\chi_{\rm dof}^2$ \\
\hline
\ sol--A  \ & \ $1606\pm5$ \  &  $65\pm11$  & \ $77\pm27$ & $-28\pm9$   &
$-5\pm4$     &  \text{\scriptsize{$\mathit{-11\pm18}$}}   & \ $27\pm14$ \  &
$-10\pm3$    & \text{\scriptsize{$\mathit{10\pm3}$}}    &   $0.00$   \\
sol--B  & $1603\pm9$ & $75\pm8 $ & $81\pm28$  & $-3.6\pm5$  & $-0.4\pm3.2$ & \
\text{\scriptsize{$\mathit{23\pm14}$ }} & $27\pm12$  & \ $-4\pm6$ &
\text{\scriptsize{$\mathit{4\pm6}$}} & $0.00 $ \\
OGE    & $1607\pm6$ & $74\pm9$  & $80\pm28$  & \ $-0.4\pm3$ \
&\text{\scriptsize{$\mathit{0.4\pm3}$}} & \text{\scriptsize{$\mathit{30\pm17}$}} & $30\pm14$ & $-5\pm4$ & \text{\scriptsize{$\mathit{5\pm4}$}} & $0.22$  \\
OME    & $1605\pm4$ & $63\pm12$ & \text{\scriptsize{$\mathit{63\pm12}$}}  & \text{\scriptsize{$\mathit{-26\pm16}$}}  & $-5\pm4$ &\text{\scriptsize{$\mathit{-20\pm10}$}}& $15\pm6$ & $-8\pm2$ & \text{\scriptsize{$\mathit{8\pm2}$}} & $0.53$ \\
\hline
\hline
\end{tabular}
\caption{The sol--A and sol--B fits of the general interaction have seven
free parameters. The fits of the OGE and OME interactions have six and five
parameters, respectively. The smaller numbers in italic indicate
which parameters have not been fitted and were obtained instead from the constraints.
All parameters are given in MeV.}
\label{tab:fits}
\end{table}

\begin{table}[h]
\begin{tabular}{ccccc} 
\hline
\hline
 & sol--A & sol--B  &  OGE  & OME \\
\hline
$\theta_1$  \quad \quad  & \ $0.80\pm0.32$ \ & \ $-0.04\pm0.68$ \ &  \ $-0.04\pm0.19$ \ & $0.78\pm0.28$ \\
$\theta_3$  \quad\quad   &   $0.01\pm0.21$   &   $-0.23\pm0.17$   &    $-0.29\pm0.17$   & $0.11\pm0.13$ \\
\hline
\hline
\end{tabular} 
\caption{Mixing angles (in radians) for the two solutions (sol--A and sol--B) of
the general OGE+OME interaction and for the OGE
and OME model interactions.}
\label{table4} 
\end{table}

\begin{figure}[tb]
\includegraphics[width=7cm,keepaspectratio,angle=0,clip]{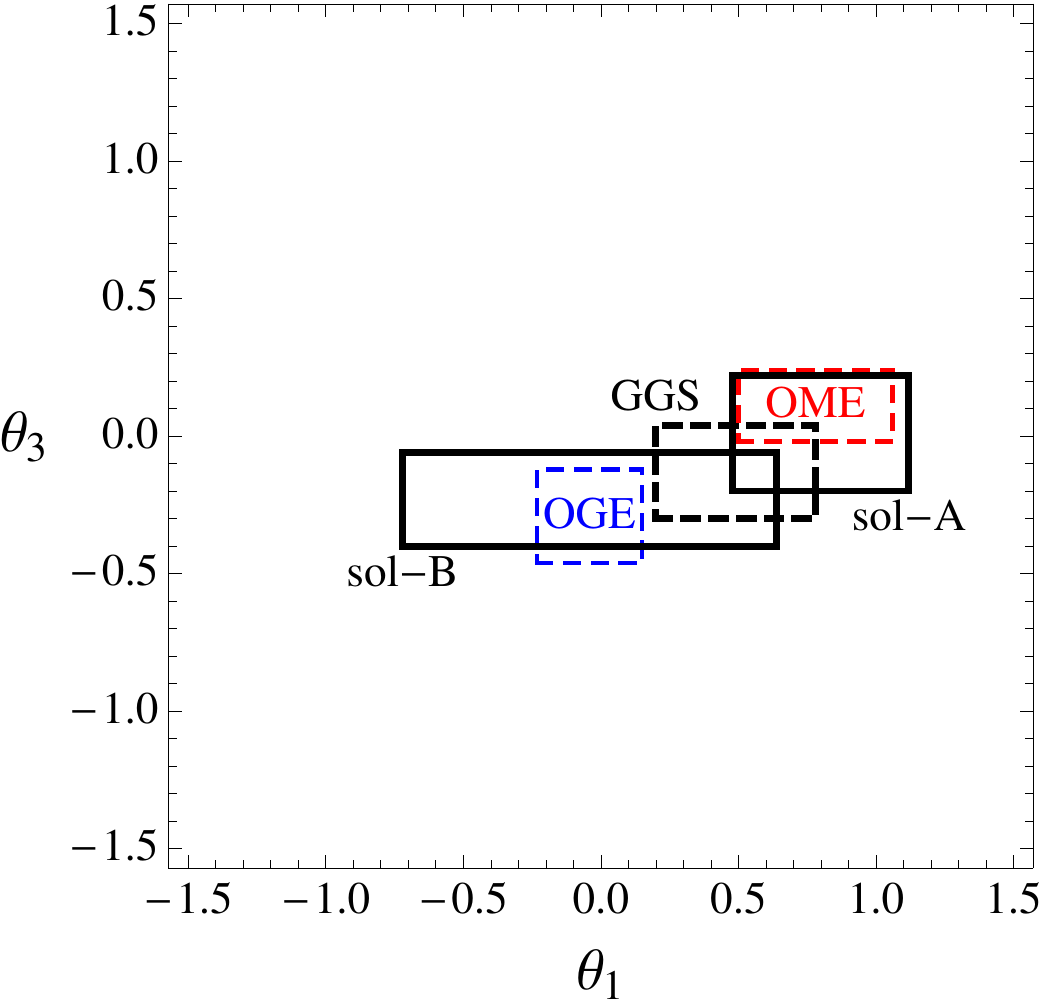}
	\caption{Mixing angles for sol--A, sol--B, the OME and OGE model interactions as in Table~\ref{table4} 
and the GGS best 
global fit values of Ref.~\cite{deUrreta:2013koa} } 
\label{fig:summary}
\end{figure}

\begin{figure}[tb]
\includegraphics[width=15cm,keepaspectratio,angle=0,clip]{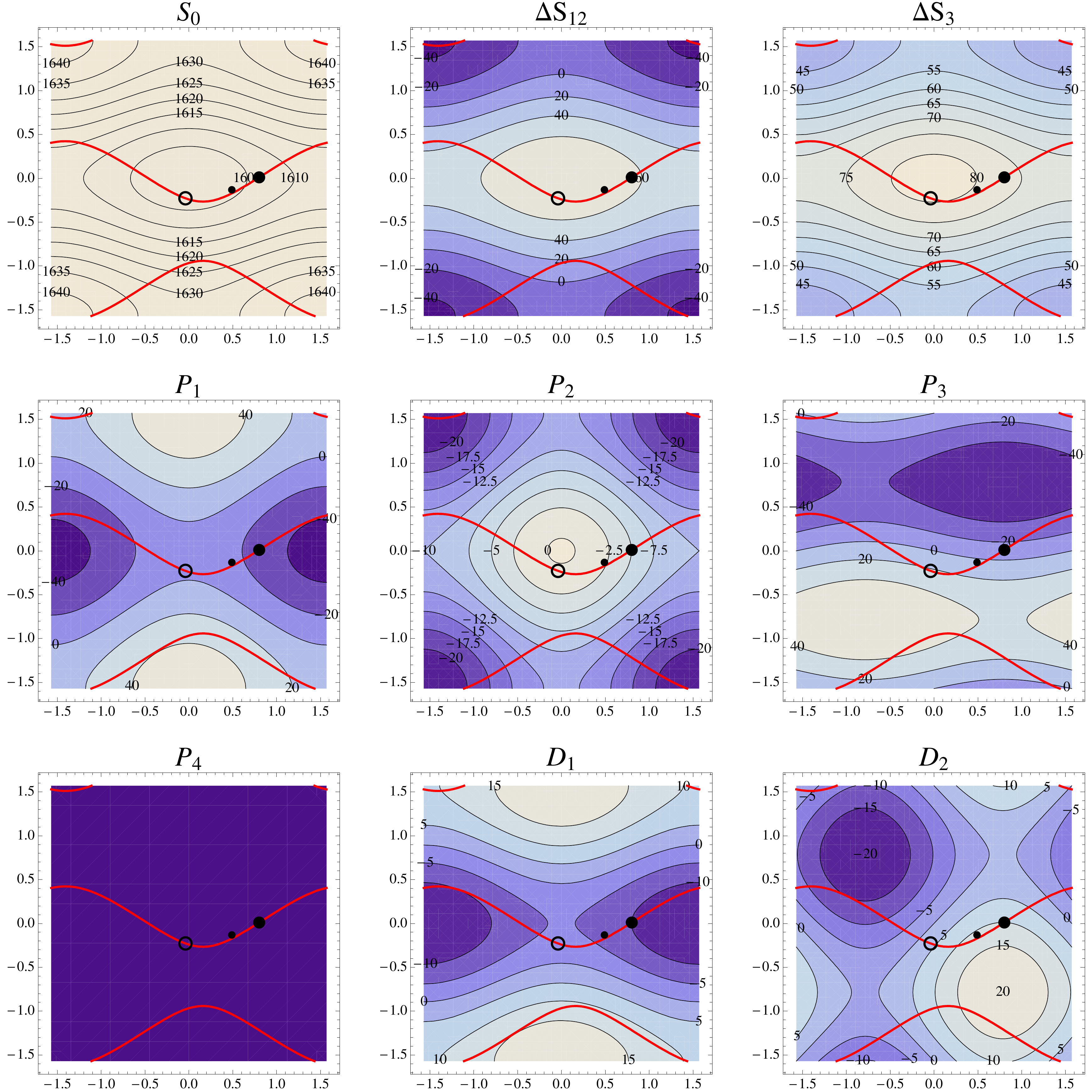}
\caption{ Matrix elements as a function of the mixing angles
$(\theta_1,\theta_3)$. We also show the mixing angles for sol--A (black dot) and 
sol--B (circle).  The smaller point in between corresponds to the GGS best fit
angles. The red curve shows the correlation between mixing angles given by
relation R1, Eq.~(\ref{rel1}). } 
\label{fig:SPD}
\end{figure}

\begin{figure}[tb]
\includegraphics[width=7cm,keepaspectratio,angle=0,clip]{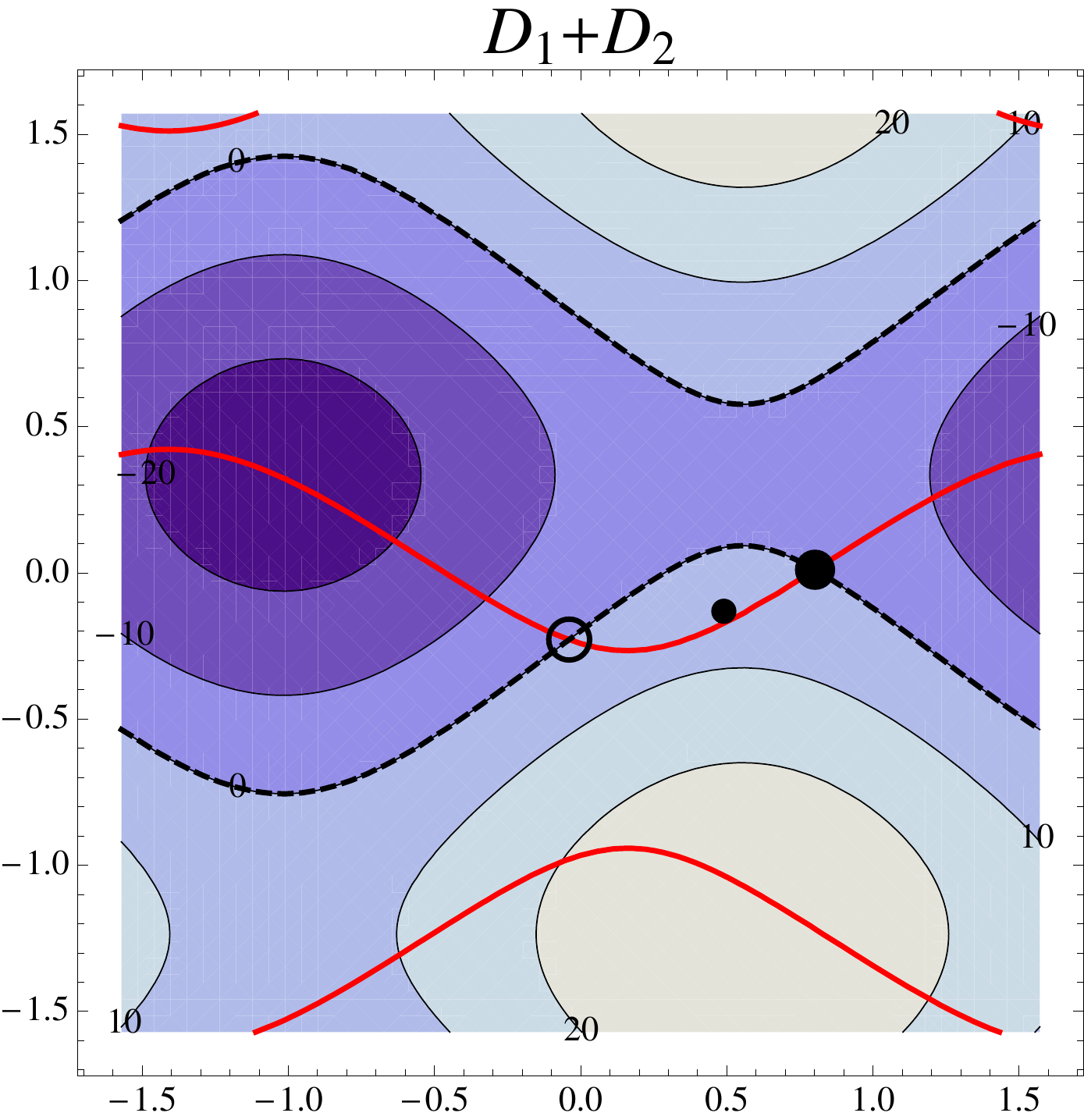}
\qquad
\qquad
\includegraphics[width=7cm,keepaspectratio,angle=0,clip]{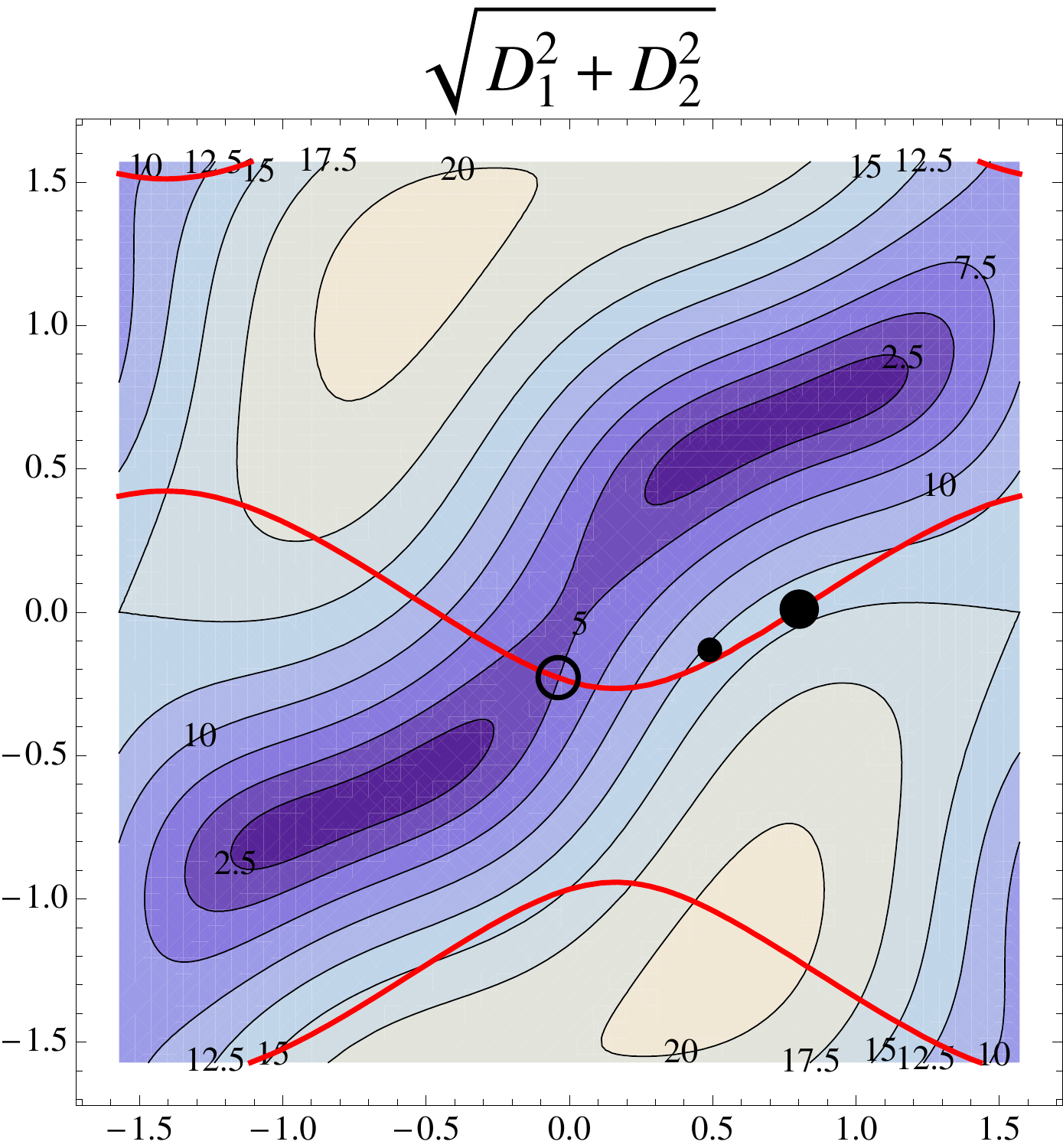}
	\caption{ The constraint $D_1+D_2$  (dashed contour in left panel) and the magnitude of the tensor interaction 
$\sqrt{D_1^2+D_2^2}$ (right panel) as a function of the mixing angles. R1, sol--A, sol--B and best 
fit value as in Fig.~\ref{fig:SPD}.   } 
\label{fig:D12}
\end{figure}

\begin{figure}[tb]
\includegraphics[width=7cm,keepaspectratio,angle=0,clip]{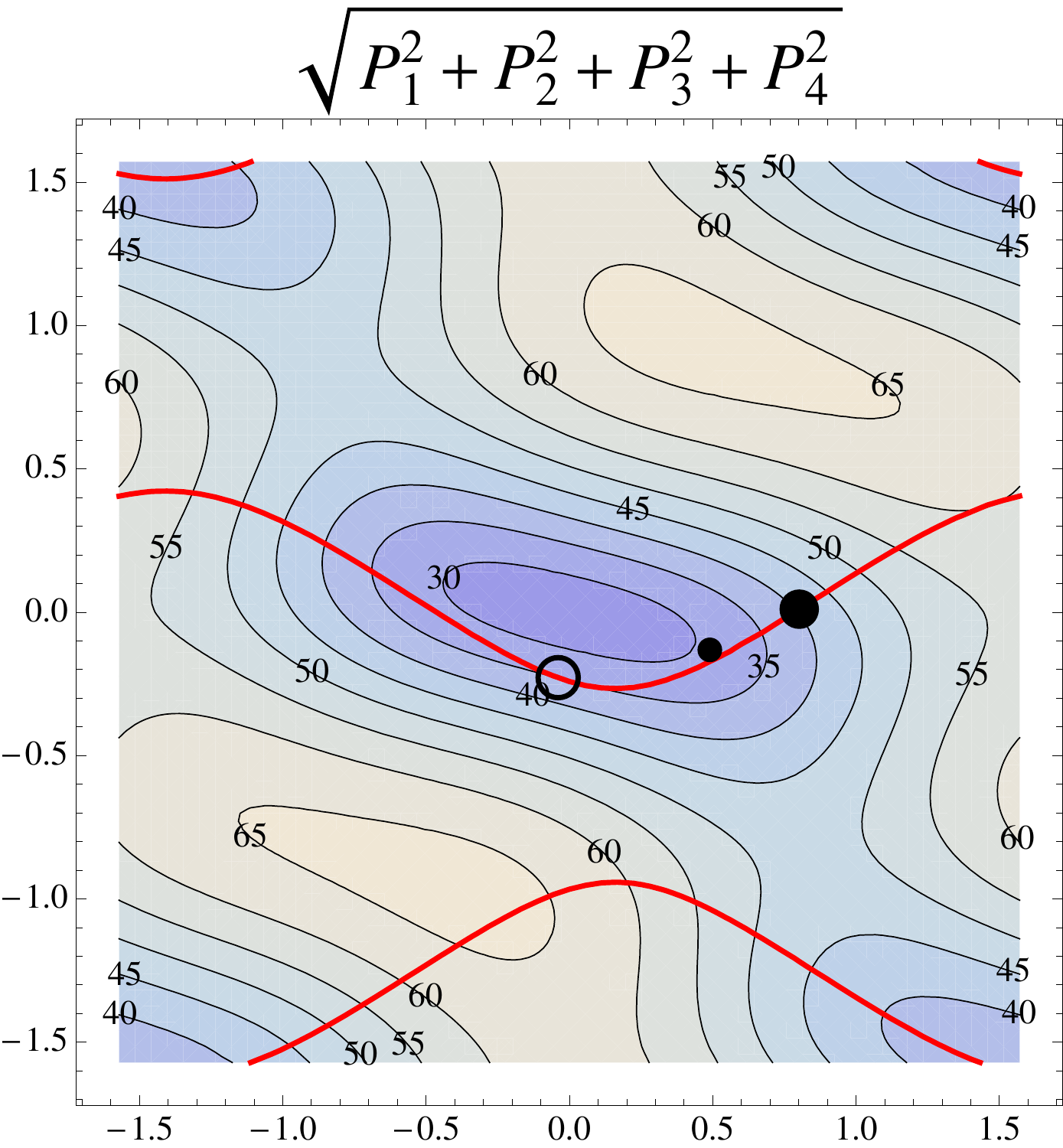}
	\caption{ The magnitude of the spin-orbit interaction 
$\sqrt{P_1^2+P_2^2+P_3^2+P_4^2}$ as a function of the mixing angles. 
R1, sol--A, sol--B and best 
fit value as in Fig.~\ref{fig:SPD}. } \label{fig:P1234}
\end{figure}

In this Section we discuss the best fits to the mass matrix of the $L=1$
excited baryons and the values we obtain for the mixing angles.
We find that the two possible solutions for the angles of the
general OGE + OME spin-flavor interaction are well
approximated by the angles obtained in the more restrictive cases of an OGE or
OME interaction. 

For the general OGE+OME form of the
quark-quark interaction, Eqs.~(\ref{vss}-\ref{vt}), we obtain
the two mass-angle relations $\rm R1, R2$, that follow from 
the two constraints,  Eq.~(\ref{relP-OGEOME}) and Eq.~(\ref{relD-OGEOME}),  leaving only  
seven parameters free,
which are fitted to the seven masses and the two mixing angles obtained as
sol--A and sol--B in the previous Section.
In the restricted model interactions, in addition to 
R1, R2 we have more relations that follow from further constraints.
In the OGE case we find
one more relation, Eq.~(\ref{relP-OGE}), leading to a six parameters fit of the seven empirical 
masses.  In the OME case there are two additional relations, Eqs.~(\ref{relS-OME},\ref{relP-OME}),  
leaving five parameters free for the fit. 
In both cases the corresponding mixing angles are
obtained diagonalizing the best fit mass matrix. The errors in the mixing angles are obtained by 
error propagation using the expressions of Appendix~\ref{massmatdiag}.

Before discussing the results of the fits, it is useful to 
redefine the parameters of the scalar part of the interaction as 
\begin{eqnarray}\label{redefS}
S_1 &=& S_0 - \Delta S_{12} \ , \qquad 
S_2 = S_0 + \Delta S_{12} \ ,  \qquad
S_3 = S_0 + \Delta S_{3}\ . 
\end{eqnarray}
In this way we can identify $S_0$ as the spin-independent contribution of the unit operator
and separate it explicitly, as that contribution is at the much larger energy
scale of the average mass of the multiplet and does not contribute to the mass
splittings. 

The results of the fits are summarized in Table~\ref{tab:fits}, 
where the small numbers in italic indicate which parameters have 
not been fitted and were determined instead from the constraints that 
express them in terms of the free parameters.
Within our minimal assumptions for the form of the quark-quark interactions,
from the fits we obtain the typical sizes of $80 \
{\rm MeV}, 30 \ {\rm MeV}$ and $ 10 \ {\rm MeV} $ for the $\ell = 0,1,2 $
components of the mass operator, respectively, with an  average mass $S_0
\approx 1600 \ {\rm MeV}$. 
The predictions for the OGE and OME mass spectrum are 
included in Table~\ref{tab:exp}.
Both model interactions are capable of perfectly fitting the data 
without any noticeable tension. 

Our results for the mixing angles are summarized in Table~\ref{table4} and
Fig.~\ref{fig:summary}, where they are also compared with the best fit angles
of the global fit of Ref.~\cite{deUrreta:2013koa}, labeled here as GGS, that uses the
$1/N_c$ hierarchy and also includes data on strong and electromagnetic decays.
The two general solutions sol--A and sol--B can be associated to two different
type of quark-quark interactions, labeled as OGE and OME.  Within the present 
experimental uncertainties the global fit to
the angles cannot distinguish among these two different possibilities.

The two solutions sol--A and sol--B differ in the details of their 
spin-orbit and tensor interactions. In Fig.~\ref{fig:SPD} we show how the
$S,P,D$ parameterization of the mass matrix depends on the mixing angles. We
include R1 (red curve), sol--A (black dot), sol--B (circle) and the GGS best
fit (small black dot) in all contour plots, serving as reference points.
Fig.~\ref{fig:SPD} shows that the $S,P,D$ parameters are smooth functions of the angles,
in contrast to the $c_i$ operator coefficients  of the $1/N_c$ studies that
will be discussed in the next Section. In particular, $P_4$ is independent of
the mixing angles. 
On the left panel of Fig.~\ref{fig:D12} we show the $D_1+D_2$ constraint,
Eq.~(\ref{relD-OGEOME}), which is a slowly varying function of the angles and is also
approximately well satisfied by the best fit values.   On the right panel we
plot  $\sqrt{D_1^2+D_2^2}$ to obtain a picture of the magnitude of the tensor
interaction.  We can see that sol--B is close to a region of small tensor
interaction, while sol--A moves away from that region. In  Fig.~\ref{fig:P1234}
we plot $\sqrt{P_1^2+P_2^2+P_3^2+P_4^2}$ and we see that both sol--A and sol--B
have spin-orbit matrix elements of similar strength.

From the fits (see Table~\ref{tab:fits}) we observe that $\Delta S_{12} \approx \Delta S_{3}$ (i.e. $S_2 \approx S_3$), 
which is the telltale sign of a contact
interaction in the flavor independent spin-spin  part, as discussed at the end of 
Sec.~\ref{sec:marel} and in Ref.~\cite{Pirjol:2010th}. This conclusion holds regardless of the 
spin-isospin structure of the interaction.
 
The values obtained for the fitted parameters can be understood qualitatively noting that there are
some special mass combinations that are independent of the angles. We will discuss them next.

In the case of $P_4$ and $\Delta S_3$, they are given by the splitting
and the spin averaged mass of the $\Delta_{3/2},\Delta_{1/2}$ as follows:
\begin{eqnarray}\label{P4}
P_4 &=& (\Delta_{3/2} - \Delta_{1/2})/3 \ ,  \\
S_3 &=& (2 \Delta_{3/2} +  \Delta_{1/2})/3 \ , 
\end{eqnarray}
giving $P_4 \approx 27  \ {\rm MeV} $,  $\Delta S_3 = S_3-S_0 \approx 78 \  {\rm MeV}$ using the 
empirical masses of Table~\ref{tab:exp}.

Taking the trace of the mass matrix for $J=1/2$ and $J=3/2$ we obtain the simple mass relation 
\begin{eqnarray}\label{traceRel}
N_{1/2} + N'_{1/2} &=& N_{3/2} + N'_{3/2} \ , 
\end{eqnarray}
which is independent of the mixing angles and is observed to be well satisfied by the empirical masses as  
$3193 \, \pm \, 16 \ {\rm MeV} = 3220 \, \pm \, 50  \ {\rm MeV}$. It should therefore hold for all 
fits that accurately describe the spectrum, which for the matrix elements implies 
the constraint
\begin{eqnarray}\label{traceRelPar1}
3 D_1 &=& P_1-P_2 \ .  
\end{eqnarray}
This constraint is satisfied in different ways for the different fits we
considered. 
In sol--B and the OGE fit both $P_1$ and $P_2$ are small, implying a small value of the tensor matrix element
$D_1$. 
In sol--A and the OME fit $P_2$ is small, but $P_1$ is large, which gives a larger $D_1$
(see right panel of Fig.~\ref{fig:D12}).

Another empirical relation that is well satisfied is
\begin{eqnarray}\label{N52}
N_{5/2} &=& (2 \Delta_{3/2} +  \Delta_{1/2})/3 \ , 
\end{eqnarray}
as seen by inserting the empirical values for the masses: $1675 \pm 5  \ {\rm MeV} = 1683 \pm 28  \ {\rm MeV} $. 
For the matrix elements this implies 
\begin{eqnarray}\label{traceRelPar2}
S_2 - S_3&=& 3 P_2 - D_1 \ .  
\end{eqnarray}
Using $S_2=S_3$, which is satisfied within errors by all the fits,   we get $D_1 = 3 P_2$ as an approximate relation
that explains  
the smallness of the spin-orbit matrix element $P_2$ as observed in all our fits 
correlating it with the typical size of the tensor matrix elements $D_{1,2}$.

\section{Relation to $1/N_c$ studies}
\label{sec:Nc}
In this Section we translate the constraints on matrix elements that we
discussed in the previous Sections to constraints on the coefficients $c_i$ to
gain further insight in their physical meaning, connecting in this way the QCD
based approach of the $1/N_c$ studies to a dynamical calculation in a quark
model.

\begin{figure}[htb]
\includegraphics[width=15cm,keepaspectratio,angle=0,clip]{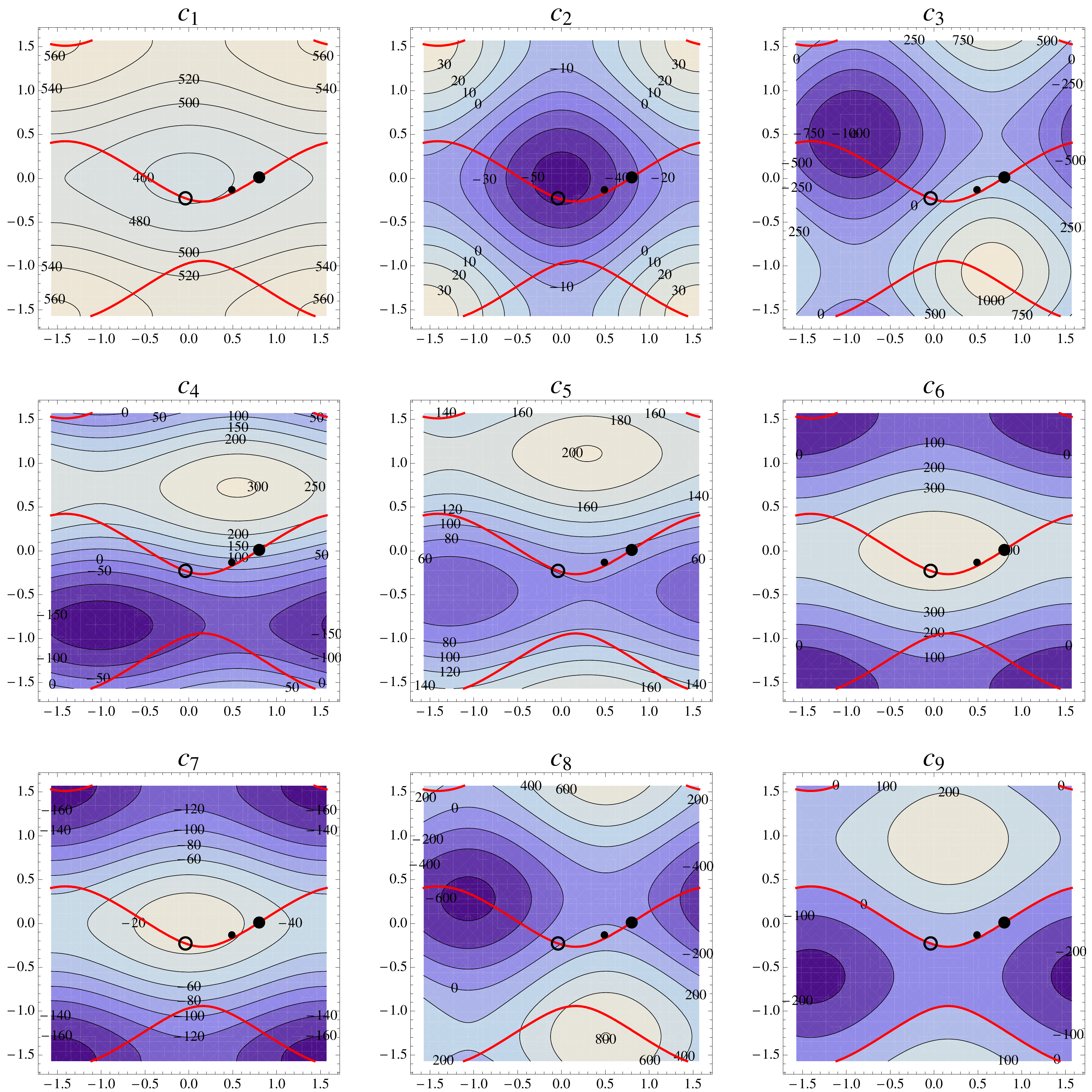}

\caption{Operator coefficients as a function of the mixing angles
$(\theta_1,\theta_3)$. We also show the mixing angles for sol--A (black dot)
and sol--B (circle).  The smaller point in between corresponds to the best fit
angles. The red curve shows the correlation between mixing angles given by
relation R1, Eq.~(\ref{rel1}). }
 
\label{fig:c1to9} \end{figure}

By choosing a basis of nine independent $1/N_c$ operators we can
reexpress the matrix elements $S_{1,2,3}, P_{1,2,3,4}, D_{1,2}$ in terms of the
operator coefficients $c_i \ (i=1,9)$. For the explicit relation between both
parametrizations see Eqs.~(\ref{b1c1})-(\ref{b1c9}) in Appendix~\ref{cbasis}.
For the nine operator coefficients $c_{1 \dots 9}$ Fig.~\ref{fig:c1to9} gives
the landscape on the mixing angles plane. We observe that all coefficients
depend on the angles and $c_3$ and $c_8$ have the steepest slopes and are very
sensitive to the mixing angle values, in contrast to the $S, P, D$ parameters (see
Fig.~\ref{fig:SPD}) were all the matrix elements were slowly varying functions
of the angles $\theta_1, \theta_3$.

In Table~\ref{table6} we present the values we obtain for the coefficients $c_i$ that correspond to the
different cases we discussed: sol--A and sol--B for a general interaction of
OGE + OME type, and the restricted  OGE and OME models.  They are shown
together with two recent determinations,
Refs.~\cite{deUrreta:2013koa,Fernando:2014dna}, labeled as FG and GGS 
respectively.  The coefficients that
correspond to the GGS global fit of Ref.~\cite{deUrreta:2013koa} are presented in a cartesian
basis using the conversion factors that can be found in
Ref.~\cite{Fernando:2014dna}.  
The small numbers in italic indicate which coefficients have 
not been fitted and were determined instead from constraints that 
will be discussed below.

We will compare in the following with  the values of the $c_i$ obtained from 
the GGS global fit of Ref.~\cite{deUrreta:2013koa}, where the hierarchy of
$1/N_c$ operators was used, and the data on strong and electromagnetic decays
was also taken into account.

The first general relation,  Eq.~(\ref{relP-OGEOME}),  corresponds to $c_9=0$ and
is taken as a constraint in Refs.~\cite{deUrreta:2013koa,Fernando:2014dna} 
as it corresponds to neglecting $O_9$, a higher order
operator in the $1/N_c$ expansion.  The second general relation,
Eq.~(\ref{relD-OGEOME}),  corresponds to $c_3+ 4 \, c_8=0$.  This evaluates to
$104 \pm 50 \ {\rm MeV}=0$ for GGS, which at first sight seems to be violated
badly.  However, the relation to $D_1+D_2$ is given by Eq.~(\ref{cR2}), which
for the $O_{1\dots9}$ basis evaluates to $c_3+ 4 \, c_8= 144 (D_1+D_2)$. 
The large numerical factor in this expression finally gives $D_1 + D_2 = 0.7 \pm 0.3 \
{\rm MeV}$,  which actually amounts to a small violation of the angle
correlation imposed by $D_1+D_2=0$, as $D_1+D_2$ is a slowly varying function 
(see left panel of Fig.~\ref{fig:D12}).

Restricting the general OGE+OME interaction to the OGE, OME model interactions
we obtain from the OGE relation Eq.~(\ref{relP-OGE}) the equivalent $c_4=0$
constraint, which is satisfied within errors as $38 \pm 39 \ {\rm MeV} = 0$.   The first of
the OME relations, Eq.~(\ref{relS-OME}), implies $c_7=0$  which is satisfied as
$4\pm13 \ \rm MeV = 0$. The second OME relation, Eq.~(\ref{relP-OME}), implies
$ 9 \,c_4 = - 12 \, c_2 + 16 \, c_5$ and is also satisfied as $ 342 \pm 351 \
{\rm MeV} = 1232 \pm 314  \ {\rm MeV}$.

We see that the general relations, as well as the more restricted OGE and OME
relations are compatible with the GGS global fit values for the coefficients. 
This is consistent with
what was observed for the mixing angles, as discussed in Sec.~\ref{sec:fits}
and summarized in Fig.~\ref{fig:summary}. 

If the general  OGE+OME interaction with a contact interaction for the 
flavor independent spin-spin part is considered, then $S_2=S_3$ also implies $c_7=0$. This is well
satisfied in all fits,  giving a clear indication of a short range interaction
in the flavor independent hyperfine part of the potential, one of the main conclusions of
Ref.~\cite{Pirjol:2010th}.

Finally we consider the constraints imposed by the two empirical relations,
Eq.~(\ref{traceRel}) and Eq.~(\ref{N52}).  The trace relation,
Eq.~(\ref{traceRel})  gives  $3 \,c_3 + 8 \,c_5 = -12 \, c_2 + 9 \, c_4 $ which
is satisfied within errors and evaluates to $ 760 \pm 322 \ {\rm MeV} = 642 \pm
427  \ {\rm MeV}$.  The $N_{5/2}$ relation, Eq.~(\ref{N52}), gives $-3 \,c_2 +
\frac16 \,c_3 = 2 \, c_5 + \frac13 \, c_8 $ which is also satisfied within errors: $
88 \pm 60 \ {\rm MeV} = 119 \pm 74  \ {\rm MeV}$. This is consistent with the
good description of the spectrum obtained in the $1/N_c$ studies.

All these correlations among operator coefficients are not obvious when 
blindly performing a fit to the mass spectrum. The matching of a general quark model to  the effective spin-flavor 
operators used in the $1/N_c$ expansion uncovers them and gives an insight into 
their dynamical meaning.

\begin{table}
\begin{tabular}{ccccccc} 
\hline
\hline
\ \ \ \quad \quad  \   \ \  	& sol--A &  sol--B	& OGE 			& OME		 	& FG  & GGS		\\
\hline
$c_1$            	& $467\pm10$ 	& $458\pm9$		& $460\pm9$ 		& $472\pm12$ 		& $463\pm2$     	& $497\pm5$		\\
$c_2$            	& $-33\pm28$ 	& $-52\pm22$		& $-61\pm22$ 		& $-10\pm8$ 		& $-36\pm12$    	& $-24\pm20$		\\
$c_3$            	& $241\pm76$ 	& $85\pm132$		& $116\pm104$	& $200\pm52$ 	& $313\pm69$    	& $96\pm42$		\\
$c_4$            	& $132\pm48$ 	& $16\pm27$		& \text{\scriptsize{$\mathit{0}$}}	& \text{\scriptsize{$\mathit{124\pm41}$}} 	& $65\pm31$     	& $38\pm39$		\\
$c_5$            	& $95\pm49$  	& $81\pm42$		& $88\pm47$ 		& $62\pm29$ 		& $71\pm18$     	& $59\pm37$		\\
$c_6$            	& $408\pm60$ 	& $460\pm50$		& $456\pm51$ 		& $367\pm75$ 		& $443\pm10$    	& $283\pm36$		\\
$c_7$            	& $-27\pm64$ 	& $-12\pm65$		& $-11\pm65$ 		& \text{\scriptsize{$\mathit{0}$}}	& $-20\pm31$    	& $4\pm13$		\\
$c_8$            	& \text{\scriptsize{$\mathit{-60\pm20}$}}  	& \text{\scriptsize{$\mathit{-21\pm33}$}} 		& \text{\scriptsize{$\mathit{-29\pm26}$}} 		& \text{\scriptsize{$\mathit{-50\pm13}$}} 		&\text{\scriptsize{$\mathit{0}$}}	& $2\pm7$		\\
$c_9$            	& \text{\scriptsize{$\mathit{0}$}}	
                    & \text{\scriptsize{$\mathit{0}$}}		
                    & \text{\scriptsize{$\mathit{0}$}}	
                    & \text{\scriptsize{$\mathit{0}$}}
                    & \text{\scriptsize{$\mathit{0}$}}	           	
                    & \text{\scriptsize{$\mathit{0}$}}			\\
$\theta_1$  	& $0.80\pm0.32$ & $-0.04\pm0.68$ 	& $-0.04\pm0.19$		& $0.78\pm0.28$ 	& $0.52\pm0.13$ 	& ${0.49\pm0.29}$	\\
$\theta_3$  	& $0.01\pm0.21$ & $-0.23\pm0.17$ 	& $-0.29\pm0.17$		& $0.11\pm0.13$		& $-0.12\pm0.09$ 	& ${-0.13\pm0.17}$	\\
\hline
\hline
\end{tabular} 
\caption{Best fits for the operator coefficients (in MeV) and mixing angles (in radians). 
The last two columns show the results of Refs.~\cite{deUrreta:2013koa,Fernando:2014dna}, respectively.
The smaller numbers in italic indicate
which coefficients have not been fitted and were obtained from constraints.
}
\label{table6} 
\end{table}

\section{Conclusions}
\label{sec:conclusions}

We showed that for a large class of non-relativistic quark models, as defined
by Eqs.~(\ref{Hamiltonian})-(\ref{vt}),  with flavor-dependent and
flavor-independent  quark-quark interactions, the observed mass spectrum of
the non-strange $L=1$ excited baryons sets stringent constraints on the two
mixing angles of the $J=1/2,3/2$ states.  Noteworthy, these constraints are
independent of the detailed shape of the spin-spin,
spin-orbit or tensor components of the effective quark-quark interaction.
This is made manifest by expressing them as the solutions (sol--A and sol--B)
of the two parameter-free relations R1, R2 between empirical masses and mixing
angles, shown in Eq.~(\ref{rel1}) and Eq.~(\ref{rel2}).

Restricting the form of the interactions to flavor-independent (OGE) or
flavor-dependent (OME) terms only, we still obtain to a good approximation
each of the two solutions we have in the most general case (OGE+OME).  In
the OGE case we obtain one additional correlation R3, Eq.~(\ref{rel4}), that is
compatible with sol--B, while in the OME case we find two new parameter-free
relations R4, R5, see Eqs.~(\ref{rel5},\ref{rel6}), that are compatible with
sol--A.  

Only the first of the two OGE+OME relations (R1) was previously obtained in
\cite{Pirjol:2008gd} by considering the most general two-body interactions. 
The angle correlation it implies was previously discussed 
in the  $1/N_c$ analysis of Ref.~\cite{Pirjol:2003ye} and more recently given 
in the form of R1', Eq.~(\ref{rel1p}),  by neglecting $1/N_c^2$ operators
associated to three-body forces \cite{deUrreta:2013koa,Fernando:2014dna}.  This relation alone
is not sufficient to determine the angles from the mass spectrum without
further constraints.  

In Ref.~\cite{deUrreta:2013koa} additional constraints
from the $1/N_c$ expansion and the empirical data on decays were included in a
global fit, resulting in the preferred GGS values for the mixing angles, which we use to
compare with the constraints we obtained from the matching to a generic OGE+OME
quark model. 
Within the present experimental errors we cannot exclude any of the two general
solutions  for the mixing angles, sol--A and sol--B,  each of which can be well
approximated by the restricted OME and OGE model interactions, respectively.
This situation is summarized in Fig.~\ref{fig:summary}. 

We also discussed in detail the matching of the $1/N_c$ quark operator
expansion to the non-relativistic quark model, obtaining analytic expressions
for all the operator coefficients in terms of integrals of the radial part of
the interactions, which are left unspecified in Eq.~(\ref{integrals}).  The
explicit relation between our parametrization of the mass matrix and the
$1/N_c$ basis used in Refs.~\cite{deUrreta:2013koa,Fernando:2014dna} is given
by  Eq.~(\ref{b1c1})-(\ref{b1c9}) in Appendix~\ref{cbasis}, establishing 
together with Eqs.~(\ref{SPDIUJ},\ref{integrals},\ref{eqs1}-\ref{eqd2}) the 
analytic correspondence with a large class of model interactions.

Showing explicitly how the QCD based $1/N_c$ analysis is related to a model
calculation is useful to get an insight into the dynamics that generates the
values of the fitted coefficients and mixing angles. It also uncovers
correlations among the coefficients that are otherwise unnoticed when
performing a fit and relates them to the attributes of the effective quark-quark
interaction, as discussed in detail in Sec.~\ref{sec:marel} and Sec.~\ref{sec:Nc}.   

We expect that the forthcoming results of lattice calculations, that will
provide smaller errors in the mass spectrum and the possibility to explore the
dependence of the baryon spectrum and the mixing angles as a function of the
quark masses (see Ref.~\cite{Fernando:2014dna} for a pioneer effort in this
direction), in combination with the $1/N_c$ expansion supplemented by a
dynamical picture as provided here by matching to the quark model, will complement 
each other and constitute a
solid toolbox for gaining further insight into the features of the
non-perturbative regime of the strong interactions.

\begin{acknowledgments} 
C.S. thanks Dan Pirjol for discussions, a careful reading and insightful comments on the 
manuscript.
C.S. and C.W. thank the kind hospitality of the Institut f\"ur Theoretische
Physik II, University of Bochum.  C.W. thanks discussions with J.L.~Goity and
the hospitality of the Theory Group at Jefferson Lab.  We both thank J.L.~Goity,
N.N.~Scoccola and E.G. de Urreta for a discussion on R1' in
Ref.~\cite{deUrreta:2013koa} and Rodolfo Sassot for his support and encouragement.
\end{acknowledgments}

\pagebreak  
\appendix
\section{CCGL operators}
\label{CCGLops}
Here we reproduce the list of 18 effective spin-flavor operators of Ref.~\cite{Carlson:1998vx}:
\begin{eqnarray}
 {\cal O}_1 &=& N_c \openone \ , 			\\
 {\cal O}_2 &=& l s				\ , \\
 {\cal O}_3 &=& \frac{1}{N_c} l^{(2)} gG_c	\ , \\
 {\cal O}_4 &=& ls + \frac{4}{1+N_c} ltG_c	\ , \\
 {\cal O}_5 &=& \frac{1}{N_c} lS_c		\ , \\
 {\cal O}_6 &=& \frac{1}{N_c} S_c^2		\ , \\
 {\cal O}_7 &=& \frac{1}{N_c} sS_c		\ , \\
 {\cal O}_8 &=& \frac{1}{N_c} l^{(2)}sS_c	\ , 	\\
 {\cal O}_9 &=& \frac{N_c+1}{N_c} {\cal O}_4+{\cal O}_5	+ \frac{8}{N_c^2}l^ig^{ja}\{S_c^j,G_c^{ia}\}	\ , \\
 {\cal O}_{10} &=& \frac{1}{N_c} lgT_c	\ , 	\\
 {\cal O}_{11} &=& \frac{1}{N_c} tT_c	\ , 	\\
 {\cal O}_{12} &=& \frac{1}{N_c^2}l^{(2)}t\{S_c,G_c\}	\ , \\
 {\cal O}_{13} &=& \frac{1}{N_c^2} (ls)S_c^2	\ , \\
 {\cal O}_{14} &=& \frac{1}{N_c^2} \{lS_c,sS_c\}	\ , \\
 {\cal O}_{15} &=& \frac{1}{N_c^2} (lS_c)(tT_c)	\ , \\
 {\cal O}_{16} &=& \frac{1}{N_c^2} gS_cT_c	\ , \\
 {\cal O}_{17} &=& \frac{1}{N_c^2} l^{(2)}S_cS_c	\ , \\
 {\cal O}_{18} &=& \frac{1}{N_c^2} l^{(2)}gS_cT_c \ . 
\end{eqnarray}
For $N_c=3$ these 18 operators provide an overcomplete basis for the mass
operator of non-strange mixed-symmetric orbitally excited baryons. 

\section{Operator relations}
\label{oprels}
Here we give the explicit form of the linear relations between CCGL operators
that hold for arbitrary $N_c$ on the nine-dimensional space spanned by the seven diagonal matrix
elements and two off-diagonal matrix elements that are relevant for 
the physical states at $N_c=3$,

\noindent
$\ell=0:$ 
\begin{eqnarray}
\frac{N_c+3}{2N_c^2(N_c-1)} \ O_1 - \frac{1}{(N_c-1)} \ O_6  + O_{7} + O_{11} &=& 0 \ , \\
-\frac{N_c+3}{2N_c^3(N_c-1)} \ O_1 + \frac{N_c+1}{2N_c(N_c-1)} \ O_6   + O_{16} &=& 0 \ , 
\end{eqnarray}
$\ell=1:$
\begin{eqnarray}
\frac{1}{2N_c} O_2 + \frac{N_c+1}{4N_c} \ O_4  + \frac{1}{4} \ O_5 + O_{10} &=& 0 \ , \\
-\frac{2}{N_c^2} O_2 + \frac{N_c-1}{4N_c^2} \ O_5  + \frac{N_c-1}{4N_c} \ O_9 + O_{13} &=& 0 \ , \\
- \frac{4}{N_c^2} O_2 + \frac{3(N_c+1)}{2 N_c^2} \ O_4  + \frac{4 N_c-1}{2N_c^2} \ O_5 + \frac{2
  N_c-1}{2N_c} \ O_9   + O_{14} &=& 0  \ , \\ 
\frac{2}{N_c^2} O_2 - \frac{N_c+1}{N_c^2} \ O_4  - \frac{N_c-1}{2N_c^2} \ O_5 - \frac{N_c-1}{2N_c} \ O_9
+ O_{15} &=& 0 \ , 
\end{eqnarray}
$\ell=2:$
\begin{eqnarray}
\frac{4}{N_c+1} \ O_{12}   + O_{17} &=& 0 \ , \\
-\frac{8}{N_c(N_c-1)} \ O_3 - \frac{2}{N_c-1} \ O_8   + O_{17} &=& 0 \ , \\
\frac{1}{N_c} \ O_8 + O_{18} &=& 0 \ . 
\end{eqnarray}

\section{Relation between operator coefficients and the matrix element parametrization}
\label{cbasis}

For $N_c=3$ the coefficients in $O_{\ell=0,1,2}$ are related to their matrix elements
given in Table~\ref{table1b} as:
\\

\noindent
$\ell=0:$
\begin{eqnarray}
\label{eqC1}
c_1 - \frac16 c_{11} + \frac{1}{18}c_{16} &=& \frac13 (2 S_1 - S_3) \ , \\
c_6 + \frac12 c_{11} - \frac{1}{3}c_{16} &=& -3 S_1 +S_2 + 2 S_3 \ , \\
c_7 - c_{11}  &=& 2 (S_2 - S_3) \ , 
\end{eqnarray}

\noindent
$\ell=1:$
\begin{eqnarray}
c_2 - \frac16 c_{10} + \frac{2}{9}c_{13} + \frac49 c_{14} - \frac{2}{9}c_{15} &=& -2 (2
P_2 + P_4) \ , \\
c_4 - \frac13 c_{10} - \frac{2}{3}c_{14} + \frac49 c_{15}  &=& 4 ( P_2 - P_3 + P_4) \ , \\
c_5 - \frac14 c_{10} - \frac{1}{18}c_{13} - \frac{11}{18}c_{14} + \frac19 c_{15}  &=&  P_1
- P_2 - P_3 + 4 P_4 \ , \\
c_9 - \frac16 c_{13} - \frac{5}{6}c_{14} + \frac13 c_{15}  &=& 3( P_1 + 2 P_2 - P_3 +
P_4) \ , 
\label{c9}
\end{eqnarray}

\noindent
$\ell=2:$
\begin{eqnarray}
c_8 + \frac14 c_{3} - \frac{1}{3}c_{18} - \frac43 c_{12} + \frac43 c_{17} &=& 36 (D_1 +
D_2)   \ , \label{cR2} \\
c_{3} - \frac43 c_{12} + \frac43 c_{17} &=& 24 (D_1 + 2 D_2)  \ . 
\label{eqC9}
\end{eqnarray}

Notice that the right-hand-sides of Eq.~(\ref{c9}) and Eq.~(\ref{cR2}) vanish for the OGE+OME model. 

The operators $O_1$ to $O_9$ can be chosen as an independent basis for the physical states at $N_c=3$. 
Setting $c_{10}=\dots = c_{18}=0$ we can express the coefficients $c_{1 \dots 9}$ in terms
of our mass matrix parametrization: 
\begin{eqnarray}
\label{b1c1}
c_1 &=& \frac13 (2 S_1 - S_3) \ , \\
c_2 &=& -2 (2 P_2 + P_4)  \ , \\
c_3 &=& 24 (D_1 + 2 D_2)  \ , \\
c_4 &=& 4 ( P_2 - P_3 + P_4)  \ , \\
c_5 &=&  P_1 - P_2 - P_3 + 4 P_4  \ , \\
c_6 &=& -3 S_1 +S_2 + 2 S_3  \ , \\
c_7 &=& 2 (S_2 - S_3)  \ , \\
c_8 &=& 6 (5 D_1 + 4 D_2)   \ , \\
c_9 &=& 3( P_1 + 2 P_2 - P_3 + P_4)  \ . 
\label{b1c9}
\end{eqnarray}
Inverting the general relations given in Eqs.~(\ref{eqC1})-(\ref{eqC9}) we obtain 
\begin{eqnarray}
S_0 &=& 3 c_1 + \frac12 c_{6} - \frac14 c_{11}  \ , \\
 S_1 &=& 3 c_1 + \frac13 c_6 - \frac16 c_7 - \frac16 c_{11} + \frac{1}{18} c_{16} \ , \\
 S_2 &=& 3 c_1 + \frac23 c_6 + \frac16 c_7 - \frac13 c_{11} - \frac{1}{18} c_{16} \ , \\
 S_3 &=& 3 c_1 + \frac23 c_6 - \frac13 c_7  + \frac16 c_{11} - \frac{1}{18} c_{16} \ , \\ 
P_1 &=&  \frac16 c_2 - \frac14 c_4 + \frac19 c_5 + \frac{8}{27} c_9 + \frac{1}{36} c_{10} - \frac{1}{54} c_{13} - \frac{2}{27} c_{14} - \frac{1}{27} c_{15}  \ , \\
P_2 &=& -\frac16 c_2 - \frac19 c_5 + \frac{1}{27} c_9 + \frac{1}{18} c_{10} - \frac{1}{27} c_{13} - \frac{1}{27} c_{14} + \frac{1}{27} c_{15}  \ , \\
P_3 &=& -\frac13 c_2 - \frac14 c_4 + \frac19 c_5 - \frac{1}{27} c_9 + \frac19 c_{10} - \frac{2}{27} c_{13} - \frac{1}{54} c_{14} - \frac{1}{27} c_{15}  \ , \\
P_4 &=& -\frac16 c_2 + \frac29 c_5 - \frac{2}{27} c_9 -\frac{1}{36} c_{10} - \frac{1}{27} c_{13} - \frac{4}{27} c_{14} + \frac{1}{27} c_{15}  \ , \\
D_1 &=& -\frac{1}{36} c_3 + \frac{1}{18} c_8 -\frac{1}{27} c_{12} + \frac{1}{27} c_{17} - \frac{1}{54} c_{18}  \ , \\
D_2 &=&  \frac{5}{144} c_3 - \frac{1}{36} c_8 - \frac{1}{108} c_{12} + \frac{1}{108} c_{17} + \frac{1}{108} c_{18}  \ . 
\end{eqnarray}
It is convenient to remove the contribution of the unit operator to the spin-spin matrix elements 
by defining the parameters $\Delta S_{12}, \Delta S_{3}$, see Eq.~(\ref{redefS}). Here we obtain for them
\begin{eqnarray}
\Delta S_{12} &=& \frac16 c_6 + \frac16 c_{7} -\frac{1}{12} c_{11} - \frac{1}{18} c_{16}  \ , \\
\Delta S_{3} &=& \frac16 c_6 - \frac13 c_7 + \frac{5}{12} c_{11} - \frac{1}{18} c_{16}  \ .
\end{eqnarray}

\section{Angle dependence of the operator coefficients}
\label{angledep}
With the choice of $\{O_1, \dots ,  O_9 \}$ as an independent basis we can solve $c_1 \dots c_9$ in terms of 
the physical masses and mixing angles, obtaining
\begin{eqnarray}
 c_1 &=& 
 \frac19 \left(N_{1/2}  - N'_{1/2}\right) \cos 2\theta_1
 + \frac29 \left(N_{3/2} - N'_{3/2}\right) \cos 2\theta_3
 + \frac19 \left(N_{1/2} + N'_{1/2}\right) 
\nonumber \\
& & + \frac29 \left(N_{3/2} + N'_{3/2}\right)
 - \frac19 \left(\Delta _{1/2} + 2 \Delta _{3/2}\right) \ , \\
 c_2 &=& 
 \frac16 \left(N_{1/2} - N'_{1/2}\right) \cos 2\theta_1
 + \frac{2}{15} \left(N_{3/2} - N'_{3/2}\right) \cos 2\theta_3
 - \frac16 \left(N_{1/2} + N'_{1/2}\right)
\nonumber \\
& & - \frac{2}{15} \left(N_{3/2} + N'_{3/2}\right)
 + \frac35 N_{5/2}
 + \frac23 \left(\Delta _{1/2} - \Delta _{3/2}\right) \ , \\
c_3 &=& 
\left( N'_{1/2} - N_{1/2} \right) 
\Big( \cos 2\theta_1 + 4 \sin 2\theta_1 \Big) 
- \left( N'_{3/2}- N_{3/2} \right) 
\left( \frac{8}{5} \cos 2\theta_3 + 4 \sqrt{\frac{2}{5}} \sin 2\theta_3 \right) 
\nonumber \\
& & + N'_{1/2} + N_{1/2}  
- \frac{8}{5} \left( N'_{3/2}+ N_{3/2} \right)+ \frac{6}{5} N_{5/2} \ , \\
 c_4 &=& 
 - \frac16 \left(N_{1/2} - N'_{1/2}\right) \Big( \cos 2\theta_1 + 2\sin 2\theta_1 \Big)
 - \left(N_{3/2} - N'_{3/2}\right) \left(\frac{2}{15} \cos 2\theta_3 + \frac53 \sqrt{\frac25} \sin 2\theta_3 \right)
\nonumber \\
& &
 + \frac16 \left(N_{1/2} + N'_{1/2}\right)+ \frac{2}{15} \left(N_{3/2} + N'_{3/2}\right) - \frac35 N_{5/2}
 - \frac43 \left(\Delta _{1/2} - \Delta _{3/2}\right) \ , \\
 c_5 &=&
 - \frac14 \left(N_{1/2} - N'_{1/2}\right) \left(\frac12\cos 2\theta_1 + \frac13 \sin 2\theta_1 \right)
 + \left(N_{3/2} - N'_{3/2}\right)\left( \frac15 \cos 2\theta_3 - \frac16 \sqrt{\frac52} \sin 2\theta_3 \right)
\nonumber \\
& &
 -\frac{5}{24} \left(N_{1/2} + N'_{1/2}\right)
 +\frac{2}{15} \left(N_{3/2} + N'_{3/2}\right) + \frac{3}{20} N_{5/2}
 - \frac43 \left(\Delta_{1/2} - \Delta _{3/2}\right) \ , \\
 c_6 &=&
 - \frac{7}{12} \left(N_{1/2} - N'_{1/2}\right) \cos 2\theta_1
 - \frac76 \left(N_{3/2} - N'_{3/2}\right) \cos 2\theta_3
 - \frac{5}{12} \left(N_{1/2} + N'_{1/2}\right)
\nonumber \\
& & - \frac56 \left(N_{3/2} + N'_{3/2}\right)
 + \frac12 N_{5/2}
 + \frac23 \left(\Delta_{1/2} + 2 \Delta _{3/2}\right) \ , \\
 c_7 &=& 
 - \frac16 \left(N_{1/2} - N'_{1/2}\right) \cos 2\theta_1
 - \frac13 \left(N_{3/2} - N'_{3/2}\right) \cos 2\theta_3
 + \frac16 \left(N_{1/2} + N'_{1/2}\right)
\nonumber \\
& & + \frac13 \left(N_{3/2} + N'_{3/2}\right)
 + N_{5/2}
 - \frac23 \left(\Delta_{1/2} + 2 \Delta _{3/2}\right) \ , \\
 c_8 &=& 
 - \left(N_{1/2} - N'_{1/2}\right) \left(\frac54 \cos 2\theta_1 + 2 \sin 2\theta_1 \right)
 + 2 \left(N_{3/2} - N'_{3/2}\right) \left(\cos 2\theta_3 + \sqrt{\frac25} \sin 2\theta_3 \right)
\nonumber \\
& &
  + \frac54 \left(N_{1/2} + N'_{1/2}\right)
 - 2 \left(N_{3/2} + N'_{3/2}\right) + \frac32 N_{5/2} \ , \\
 c_9 &=&
 - \frac14 \left(N_{1/2} - N'_{1/2}\right) \Big(3 \cos 2\theta_1 + \sin 2\theta_1 \Big)
 + \frac12 \left(N_{3/2} - N'_{3/2}\right) \left(\frac35 \cos 2\theta_3 - \sqrt{\frac52} \sin 2\theta_3 \right)
\nonumber \\
& &
 + \frac{7}{10} \left(N_{3/2} + N'_{3/2}\right) - \frac14 \left(N_{1/2} + N'_{1/2}\right) - \frac{9}{10} N_{5/2}
 -\Delta_{1/2} +\Delta _{3/2} \ . 
\end{eqnarray}

\section{General form of the $2 \times 2$ mass matrix}
\label{massmatdiag}

We can write the mass matrix relevant for the $J=1/2$ and $J=3/2$ states in terms of the eigenvalues
$M_1$ and $M_2$ and mixing angles as follows
\begin{eqnarray}
M &=& 
\left( 
\begin{array}{cc}
M_+ + M_- \cos 2 \theta   &  M_- \sin 2 \theta \\ 
M_- \sin 2 \theta        &  M_+ - M_- \cos 2 \theta   \\ 
\end{array} 
\right) \,, 
\end{eqnarray}
with 
\begin{eqnarray}
M_+ &=& \frac12 (M_1+M_2) =  \frac12 {\rm Tr} \, M \ ,  \\
M_- &=& \frac12 (M_1-M_2) = \sqrt{\frac14 ({\rm Tr} \, M)^2  - {\rm Det} \, M }   \ , 
\end{eqnarray}
where the mixing angles enter in the change of basis matrix $C$ as:
\begin{eqnarray}
C = 
\left( 
\begin{array}{cc}
\cos \theta   &  \sin \theta \\ 
-\sin \theta  &  \cos \theta   \\ 
\end{array} 
\right) \, &,& 
C^{-1} = 
\left( 
\begin{array}{cc}
\cos \theta   &  - \sin \theta \\ 
\sin \theta  &  \cos \theta   \\ 
\end{array} 
\right) \,. 
\end{eqnarray}
We obtain the physical masses as
\begin{eqnarray}
C.M.C^{-1} &=& 
\left( 
\begin{array}{cc}
M_1   &  0     \\ 
0     &  M_2   \\ 
\end{array} 
\right) \,.
\end{eqnarray}

\section{Mass operator, mixing angles and operator expansion matching for the Isgur-Karl model}
\label{IKV}

The Isgur-Karl model \cite{Isgur:1977ef}
is defined by the quark Hamiltonian
\begin{eqnarray}
{\cal H}_{IK} = H_0 + {\cal H}_{\rm hyp}  \,, 
\end{eqnarray}
where
$H_0$ contains the confining potential and kinetic terms of the quark
fields, and is symmetric under spin and isospin. The hyperfine
interaction ${\cal H}_{\rm hyp}$ is given by 
\begin{eqnarray}\label{HIK} 
{\cal H}_{\rm hyp} = A \sum_{i<j}\Big[ \frac{8\pi}{3} \vec s_i \cdot \vec s_j
\delta^{(3)}(\vec r_{ij}) + \frac{1}{r_{ij}^3} (3\vec s_i \cdot \hat r_{ij} \
\vec s_j \cdot \hat r_{ij} - \vec s_i\cdot \vec s_j) \Big] \,, 
\end{eqnarray} 
where $A$ determines the strength of the interaction 
\footnote{In Ref.~\cite{Isgur:1977ef} A is taken as $A=\frac{2 \alpha_S}{3 m^2}$.}, and
$\vec r_{ij} = \vec r_i - \vec r_j$ is the distance between quarks
$i,j$.  The first term is a contact spin-spin interaction, and the second
describes a tensor interaction between two dipoles. Both terms are flavor independent.  
This interaction
Hamiltonian is an approximation to the gluon-exchange interaction,
neglecting the spin-orbit terms. 
The entire spectroscopy of 
the $L=1$ baryons is fixed by one
single constant $\delta$,  defined as 
$\delta = A\frac{2 \alpha^3}{\sqrt{2 \pi}}\simeq 300 \ \text{MeV}$, along with an overall additive constant $m_0$,
 the average mass for the multiplet $m_0 \simeq 1610 \ \rm MeV$, 
and
the model is very predictive. The explicit mass matrix is given by
\begin{eqnarray}\label{IKfirst} 
M_{1/2} &=& m_0 + \frac14 \delta \left(
\begin{array}{cc} 
-1 &  -1 \\ 
-1 &   0\\ 
\end{array} 
\right) \,, \\ 
M_{3/2} &=& m_0 + \frac14 \delta \left( 
\begin{array}{cc} 
-1 & \frac{1}{\sqrt{10}} \\ 
\frac{1}{\sqrt{10}} & \frac95 \\ 
\end{array}
\right) \,, \\ 
M_{5/2} &=& m_0  + \frac15 \delta \,,  \\ 
\Delta_{1/2}
&=& \Delta_{3/2} = m_0  + \frac14 \delta \,. 
\label{IKlast} 
\end{eqnarray}

The mixing angles are independent of the hadron masses, and are given by
\begin{eqnarray}\label{mixharm} 
\theta^{IK}_{1} = \arctan (\frac12 (\sqrt5 -
1)) = 0.55  \,,\qquad \theta^{IK}_{3} = \arctan
(-\frac{\sqrt{10}}{14+\sqrt{206}}) =  -0.11  \,.  
\end{eqnarray}

The matching performed in Ref.~\cite{Galeta:2009pn} to the $1/N_c$
operators obtained
$ 
c_1 = \frac13 m_0 - \frac14 \delta = 462 \ {\rm MeV}, 
c_6 = \frac32 \delta= 450 \ {\rm MeV}, 
c_8 = -\frac65 \delta= -360 \ {\rm MeV}, 
c_{17}=\frac{9}{10}\delta= 270 \ {\rm MeV}, 
$
all the other coefficients being zero.
The corresponding $\chi^2=33$ is large because, due to the absence of spin-orbit forces, the IK 
model fails to describe the splitting of the $\Delta_{1/2}(1620)$ and $\Delta_{3/2}(1700)$, 
and the mass of the $N_{1/2}(1535)$ comes out 
to low. This explains why the  $(\theta^{IK}_{1}, \theta^{IK}_{3})$ fall outside the OGE 
rectangle of our Fig.~\ref{fig-r1r2r3s1}, which was obtained from a fit with a much smaller 
value of $\chi^2$.
 
Evaluating the radial integrals, Eq.~(\ref{integrals}), we obtain for our matrix elements parametrization
$
S_1 = m_0 - \frac14 \delta = 1535 \ {\rm MeV}, 
S_2 =  
S_3 = m_0 + \frac14 \delta = 1685 \ {\rm MeV},
P_1=P_2=P_3=P_4=0, 
D_1=- 
D_2=-\frac{1}{20}\delta = -15 \ {\rm MeV} . 
$
In terms of the redefined spin-spin parameters, see Eq.~(\ref{redefS}), 
$
S_0 = m_0 = 1610 \ {\rm MeV}  , 
\Delta S_{12} = \Delta S_{3} = \frac14 \delta = 75 \ {\rm MeV} .
$
In this way the IK model provides a simple analytical check of our general expressions.

\end{document}